\documentclass[conference]{IEEEtran}

\usepackage{graphics, epsfig, psfrag, subfigure, graphicx, epstopdf}
\usepackage{algorithm, algorithmic, calc, color, multicol}
\usepackage[cmex10]{amsmath}
\usepackage{amssymb, amsfonts, stmaryrd}
\usepackage[nospace]{cite}
\usepackage{multicol, balance, bbm, setspace, url, color,verbatim, xspace}

\usepackage{url}

\newcommand {\etal}{\emph{et al.}\xspace}
\newcommand {\ie}{\emph{i.e.,~}}
\newcommand {\eg}{\emph{e.g.,~}}

\newcommand{\lastfm}{{\tt Last.fm}\xspace}
\newcommand{\facebook}{{\tt Facebook}\xspace}
\newcommand{\myspace}{{\tt Myspace}\xspace}
\newcommand{\qqq}{{\tt QQ}\xspace}
\newcommand{\orkut}{{\tt Orkut}\xspace}
\newcommand{\twitter}{{\tt Twitter}\xspace}
\newcommand{\flickr}{{\tt Flickr}\xspace}
\newcommand{\livejournal}{{\tt LiveJournal}\xspace}
\newcommand{\youtube}{{\tt YouTube}\xspace}
\newcommand{\cyworld}{{\tt Cyworld}\xspace}
\newcommand{\daum}{{\tt Daum}\xspace}

\newcommand{\eq}{\!\!=\!}

\newcommand{\Prob}{\mathbb{P}}

\newcommand{\Var}{\mathbb{V}}

\newcommand{\eqn}[1]{Eq.(#1)}

\newcommand{\red}[1]{[\textcolor{red}{\textit{#1}}]}

\begin{document}


\title{A Walk in Facebook: Uniform Sampling of Users in Online Social Networks}

\author{Minas Gjoka\\
CalIT2\\
UC Irvine\\
{\em mgjoka@uci.edu}\\
\and
Maciej Kurant\\
CalIT2\\
UC Irvine\\
{\em maciej.kurant@gmail.com}\\
\and
Carter T. Butts\\
Sociology Dept\\
UC Irvine\\
{\em buttsc@uci.edu}\\
\and
Athina Markopoulou\\
EECS Dept\\
UC Irvine\\
{\em athina@uci.edu}
}

%
%

\maketitle

\begin{abstract}

Our goal in this paper is to develop a practical framework for obtaining a uniform sample of users in an online social network (OSN) by
crawling its social graph.  Such a sample allows to estimate any user property and some topological properties as well.
To this end, first, we consider and compare several candidate crawling techniques. Two approaches that can produce approximately uniform samples are the Metropolis-Hasting random walk (MHRW) and a re-weighted random walk (RWRW). Both have pros and cons, which we demonstrate through a comparison to each other as well as to the ``ground truth.''
In contrast, using Breadth-First-Search (BFS) or an unadjusted Random Walk (RW) leads to substantially biased results.
Second, and in addition to offline performance assessment, we introduce online formal convergence diagnostics to assess sample quality during the data collection process.
We show how these diagnostics can be used to effectively determine when a random walk sample is of adequate size and quality.
Third, as a case study, we apply the above methods to  \facebook  and we collect the first, to the best of our knowledge, representative sample of \facebook users. We make it publicly available and employ it to characterize several key properties of \facebook.
\end{abstract}


\section{Introduction}

Online Social Networks (OSNs) have recently emerged as a new Internet ``killer-application.'' The adoption of OSNs by Internet users is off-the-charts with respect to almost every metric. In November 2010, \facebook, the most popular OSN, counted more than 500 million members; the total combined membership in the top five OSNs  (\facebook, \qqq, \myspace, \orkut, \twitter) exceeded 1 billion users. Putting this number into context, the population of OSN users is approaching 20\% of the world population and is  more than 50\% of the world's Internet users. According to Nielsen~\cite{nielsen-stats}, users worldwide currently spend over 110 billion minutes on social media sites per month, which accounts for 22\% of all time spent online, surpassing even time spent on email.
According to Alexa~\cite{alexa}, a well-known traffic analytics website, \facebook is the second most visited website on the Internet (second only to Google) with each user spending 30 minutes on average per day on the site (more than the time spent on Google). Four of the top five OSNs are also contained in Alexa's top 15 websites in regard to traffic rankings. Clearly, OSNs in general, and \facebook in particular, have become an important phenomenon on the Internet. %

OSNs are of interest to several  different communities. For example, sociologists employ them as a venue for collecting relational data and studying online human behavior. Marketers, by contrast, seek to exploit information about OSNs in the design of viral marketing strategies. %
From an engineering perspective, understanding OSNs can enable the design of better networked systems. For example, an OSN provider may want to understand the social graph and user activity in order to improve user experience by optimizing the design of their datacenters and/or data storage on the cloud~\cite{Pujol2010}; or by  providing personalized services and ads. A network provider may also want to understand the traffic  generated by activities of OSN users in order to design mechanisms, such as caching~\cite{agarwal2010social} and traffic engineering~\cite{nazir2009network}, to better serve that traffic. 

 Another potential use of OSNs is in algorithms that employ trusted or influential users, \eg to thwart unwanted communication while not impeding legitimate communication~\cite{mislove2008ostra}; to utilize social trust for collaborative spam filtering~\cite{Sirivianos-INFOCOM-11}; or to enable online personas to cost-effectively obtain credentials~\cite{sirivianos2009facetrust}. Third-party applications are also interested in OSNs in order to provide personalized services as well as to become popular.

The immense interest generated by OSNs has given rise to a number of measurement and characterization studies that attempt to provide a first step towards their understanding. Only a very small number of these studies are based on complete datasets provided by the OSN operators~\cite{Ahn-WWW-07, leskovec2008microscopic}. A few other studies have collected a complete view of specific parts of OSNs; \eg~\cite{harvard-dataset} collected the social graph of  the Harvard university network. However, the complete dataset is typically unavailable to researchers, as most OSNs are unwilling to share their company's data even in an anonymized form, primarily due to privacy concerns.

Furthermore, the large size\footnote{ \label{backofenvelope} A back-of-the-envelope calculation of the effort needed to crawl \facebook's social graph is as follows. In December 2010, \facebook advertised more than 500 million active users, each encoded by  64 bits (4 bytes) long userID, and  130 friends per user on average. Therefore, the raw topological data alone, without any node attributes, amounts to at least $500M\times130\times8~bytes \simeq 520~GBytes$.} and access limitations of most OSN services (\eg login requirements, limited view, API query limits) make it difficult or nearly impossible to fully cover the social graph of an OSN.
In many cases, HTML scraping is necessary, which increases the overhead multifold.\footnote{For the example in footnote \ref{backofenvelope}, one would have to download about $ 500M \times 230~KBytes \simeq 115~ TBytes$ of uncompressed HTML data.}  %
 Instead, it would be desirable to obtain and use a small but representative sample.
Given this, sampling techniques are essential for practical estimation of OSN properties.  While sampling can, in principle, allow precise population-level inference from a relatively small number of observations, this depends critically on the ability to draw a sample with known statistical properties.  The lack of a sampling frame (\ie a complete list of users, from which individuals can be directly sampled) for most OSNs makes principled sampling especially difficult. To elide this limitation, our work focuses on sampling methods that are based on \emph{crawling} of friendship relations - a fundamental primitive in any OSN.

Our goal in this paper is to provide a framework for obtaining an asymptotically uniform sample (or one that can be systematically reweighted to approach uniformity) of OSN users by crawling the social graph. We provide practical recommendations for appropriately implementing the framework, including: the choice of crawling technique; the use of online convergence diagnostics; and the implementation of high-performance crawlers. We then apply our framework to an important case-study - \facebook. More specifically, we make the following three contributions.

Our first contribution is the comparison of several candidate graph-crawling techniques in terms of sampling bias and efficiency. First, we consider Breadth-First-Search (BFS) - the most widely used technique for measuring OSNs~\cite{Mislove2007,Ahn-WWW-07} including \facebook~\cite{Wilson09}. 
BFS sampling is known to introduce bias towards high degree nodes, which is highly non-trivial to characterize analytically~\cite{Kurant2010, Kurant2011_JSAC_BFS} or to correct.
Second, we consider Random Walk (RW) sampling, which also leads to bias towards high degree nodes, but whose bias can be quantified by Markov Chain analysis and corrected via appropriate re-weighting (RWRW)~\cite{Salganik2004,Rasti09-RDS}.  Then, we consider the Metropolis-Hastings Random Walk (MHRW) that can directly yield a uniform stationary distribution of users. This technique has been used in the past for P2P sampling~\cite{Stutzbach2006-unbiased-p2p}, recently for a few OSNs~\cite{Rasti2008, Twitter08}, but not for \facebook.
Finally, we also collect a  uniform sample of \facebook userIDs (UNI), selected by a rejection sampling procedure from \facebook's 32-bit  ID space, which serves as our ``ground truth''. We compare all sampling methods in terms of their bias and convergence speed.  We show that MHRW and RWRW are both able to collect asymptotically uniform samples, while BFS and RW result in a significant bias in practice. We also compare the efficiency MHRW to RWRW, via analysis, simulation and experimentation and discuss their pros and cons. The former provides a sample ready to be used by non-experts, while the latter is more efficient for all practical purposes.

Our second contribution is that we introduce, for the first time in this context, the use of formal convergence diagnostics (namely Geweke and Gelman-Rubin) to assess sample quality in an online fashion. These methods (adapted from Markov Chain Monte Carlo applications) allow us to determine, in the absence of a ground truth, when a sample is adequate for use, and hence when it is safe to stop sampling.  These is a critical issue in implementation.

Our third contribution is that we apply and compare all the aforementioned techniques for the first time, to the best of our knowledge, on a large scale OSN. We use \facebook as a case study by crawling its web front-end, which is highly non-trivial due to various access limitations, and we provide guidelines for the practical implementation of high-performance crawlers. We obtain the first  representative sample of \facebook users, which we make publicly available~\cite{our-dataset}; we have received approximately 500 requests for this dataset in the last eighteen months. Finally, we use the collected datasets to characterize several key properties of \facebook, including user properties (\eg privacy settings) and topological properties (\eg the node degree distribution, clustering, and assortativity).

The structure of this paper is as follows. Section~\ref{sec:background} discusses related work. Section~\ref{sec:samplingmethodology} describes the sampling methodology,  including the assumptions and limitations, the candidate crawling techniques and the convergence diagnostics. Section~\ref{sec:fb_datacollection} describes the data collection process, including the implementation of high-performance crawlers, and the collected data sets from \facebook. Section~\ref{sec:fb_evalmethod} evaluates and compares all sampling techniques in terms of efficiency (convergence of various node properties) and quality (bias) of the obtained sample. Section~\ref{sec:fb_characterization} provides a characterization of some key Facebook properties, based on the MHRW sample.
Section~\ref{sec:conclude} concludes the paper. The appendices elaborate on the following points: (A) the uniform sample obtained via userID rejection sampling, used as ``ground truth'' in this paper; (B) the lack of temporal dynamics in \facebook, in the timescale of our crawls; and (C) a  comparison of the sampling efficiency of MHRW vs. RWRW.

\section{Related Work}
\label{sec:background}

Broadly speaking, there are two bodies of work related to this paper: (i) sampling  techniques, investigating the quality and efficiency of the sampling technique itself and (ii) characterization studies, focusing on the properties of online social networks based on the collected data. In this section, we review this related literature and place our work 
in perspective.

\subsection{Graph sampling techniques}

Graph sampling techniques, via crawling, can be roughly classified into two categories: graph traversal techniques and random walks.
In {\em graph traversal techniques}, nodes are sampled without replacement: once a node is visited, it is not visited again.
These methods differ in the order in which they visit the nodes; examples include Breadth-Search-First (BFS), Depth-First Search (DFS), Forest Fire (FF) and Snowball Sampling (SBS)~\cite{wasserman.faust}. 

BFS, in particular, is  a basic technique that has been used extensively for sampling OSNs in past research~\cite{Ahn-WWW-07, Mislove2007,  MisloveWosn08, Wilson09, Viswanath2009}. One reason for this popularity is that an
(even incomplete) BFS sample collects a full view (all nodes and edges) of some particular region in the graph. However, BFS has been shown to lead to a bias  towards high degree nodes in various artificial and real world topologies~\cite{Lee-Phys-Rev-06,snowball-bias,Ye2010}. Our work also confirms the bias of BFS when sampling Online Social Networks.
It is worth noting that BFS and its variants lead to samples that not only are biased but also do not have known statistical properties (and hence cannot in general be used to produce trustworthy estimates).
Although recent work suggests that it is possible to analytically compute and correct this bias for random graphs with certain degree distributions~\cite{Kurant2010}, 
these methods are mere heuristics under arbitrary graphs~\cite{Kurant2011_JSAC_BFS} and will fail for networks with large-scale heterogeneity (\eg block structure).  %
%

%

{\em Random walks on graphs} are a well-studied topic; see~\cite{Lovasz93} for an excellent survey. They have been used for sampling the World Wide Web (WWW)~\cite{ Henzinger2000, comparison-samplingweb}, peer-to-peer networks~\cite{Stutzbach2006-unbiased-p2p,Rasti09-RDS, Gkantsidis2004}, 
and other large graphs~\cite{Leskovec2006_sampling_from_large_graphs}. 
Similarly to traversals, random walks are typically biased towards high-degree nodes. However, the bias of random walks can be analyzed and corrected for using classical results from  Markov Chains.
If necessary, such a bias correction can be obtained during the walk itself - the resulting Metropolis-Hasting Random Walk (MHRW) described in Section~\ref{subsec:Metropolis-Hastings Random Walk (MHRW)} has been applied by Stutzbach~\etal~\cite{Stutzbach2006-unbiased-p2p} to select a representative sample of peers in the Gnutella network.
Alternatively, we can re-weight the sample after it is collected - the resulting Re-Weighted Random Walk (RWRW) described in Section~\ref{subsec:Re-Weighted Random Walk (RWRW)} has been recently compared with MHRW in the context of peer-to-peer sampling by Rasti~\etal~\cite{Rasti09-RDS}.
Further improvements or variants of random walks include random walk with jumps~\cite{Henzinger2000,Avrachenkov2010}, multiple dependent random walks~\cite{Ribeiro2010}, weighted random walks~\cite{Kurant2011_SWRW}, or multigraph sampling~\cite{Gjoka2011_multigraph_JSAC}. 

Our work is most closely related to the random walk techniques. We obtain unbiased estimators of user properties in \facebook using MHRW and RWRW and we compare the two through experiments and analysis; BFS and RW (without re-weighting) are used mainly as baselines for comparison. We complement the  crawling techniques with formal, {\em online convergence diagnostic tests} using several node properties.To the best of our knowledge, this has not been done before in measurements of such systems. The closest to formal diagnostics is the work by Latapy \etal~\cite{Latapy2008} which studies how the properties of interest evolve when the sample grows to practically detect steady state. We also implement {\em multiple parallel chains}. Multiple chains started at the same node have been recently used in~\cite{Rasti09-RDS}. In contrast,  we  start different chains from different nodes. We demonstrate that random walks, whose bias can be analyzed and corrected, are able to estimate properties of  users in OSNs remarkably well in practice. We also find that correcting for the bias at the end (RWRW), rather than during the walk (MHRW) is more efficient for all practical purposes - a finding that agrees with~\cite{Rasti09-RDS}.

In terms of application, we apply the measurement techniques to online social networks and study characteristics specific to that context. To the best of our knowledge, we are the first to obtain an unbiased sample of a large scale OSN, namely \facebook, and make it publicly available. Krishnamurthy \etal~\cite{Twitter08} ran a single Metropolis Random Walk, inspired by~\cite{Stutzbach2006-unbiased-p2p}, on Twitter as a way to verify the lack of bias in their main crawl used throughout the paper. However, the Metropolis algorithm was not the main focus of their paper and Twitter is a directed graph which requires different treatment. Parallel to our work, Rasti \etal {\cite{Rasti2008} also applied similar random walk techniques to collect unbiased samples of Friendster.

Previous work on the {\em temporal dynamics} of social networks includes~\cite{Kumar-KDD-06, Backstrom-KDD-06, sarkar, Willinger09-OSN_Research, Rasti2008}.  Kumar \etal~\cite{Kumar-KDD-06} studied the structure and evolution of Flickr and Yahoo!  from datasets provided by the OSN providers. Backstrom \etal~\cite{Backstrom-KDD-06} presented different ways in which communities in social networks grow over time and~\cite{sarkar} proposed a method for modeling relationships that change over time in a social network. Willinger \etal~\cite{Willinger09-OSN_Research} proposed a multi-scale approach  to study dynamic social graphs at a coarser level of granularity.  Rasti \etal~\cite{Rasti2008} evaluate the performance of random walks in dynamic graphs via simulations and show that there is a tradeoff between number of parallel samplers, churn and accuracy. In our work, we assume that the social graph remains static during the crawl, which we show in Appendix~B to be the case for \facebook in practice. Therefore, we do not consider dynamics, which are essential in other sampling contexts.

A unique asset of our study  is the collection of true uniform sample of OSN users through rejection sampling of userIDs (UNI), which served as  ground truth in this paper; see Section~\ref{sec:fb_uni}. We note that UNI yields a uniform sample of users regardless of the allocation policy of userIDs by the OSN, as shown in Appendix~A. 
UNI is essentially a star random node sampling scheme~\cite{Kolaczyk2009}; this is different from the induced subgraph random node sampling schemes that were evaluated in~\cite{Stumpf2005, Leskovec2006_sampling_from_large_graphs}.

\subsection{Characterization studies of OSNs}
Several papers have measured and characterized properties of OSNs.
In~\cite{krishnamurthy2009measure}, Krishnamurthy  presents a summary of the challenges that researchers face in collecting data from OSNs. In~\cite{Ahn-WWW-07}, Ahn \etal analyze three online social networks; one complete social graph of \cyworld obtained from the \cyworld provider, and two small samples from \orkut and \myspace crawled with BFS. %
 In~\cite{Mislove2007, MisloveWosn08}, Mislove \etal studied the properties of the social graph in four popular OSNs: \flickr, \livejournal, \orkut, and \youtube. Their approach was to collect the large Weakly Connected Component, also using BFS; their study shows  that OSNs are structurally different from other complex networks.

\cite{Wilson09, harvard-dataset, caltech-dataset} are related to this paper in that they also study \facebook. Wilson \etal~\cite{Wilson09}  collect and analyze social graphs and user interaction graphs in \facebook between March and May 2008. Their methodology is   what we refer to as Region-Constrained BFS: they exhaustively collect all open user profiles and their list of friends  in the largest regional networks.
Such Region-Constrained BFS  might be appropriate to study particular regions, but it does not provide \facebook-wide information, which is the goal of this paper. Furthermore,  the percentage of users in the social graph retrieved in~\cite{Wilson09} is 30\%-60\% less than the maximum possible in each network.\footnote{ More specifically, it is most likely that for the collection of the social graph, their BFS crawler does not follow users that have their ``view profile'' privacy setting closed and  ``view friends`` privacy setting open.We infer that,  by the discrepancy in the percentage of users for those settings as reported in a \facebook privacy study conducted during the same time in~\cite{Bala08Privacy} {\em i.e.,} in networks New York, London, Australia, Turkey.} Our findings show some noteworthy differences from~\cite{Wilson09}: for example,  we find larger values of the degree-dependent clustering coefficient, %
significantly higher assortativity coefficient, and a degree distribution that does not  follow a power law.
Finally, Wilson \etal~\cite{Wilson09} focus on the user interaction graph, while we focus on the friendship graph.
 \cite{harvard-dataset} and~\cite{caltech-dataset} have also made publicly available and analyzed datasets corresponding to university networks from \facebook with many annotated properties for each student. In contrast, we collect a sample of the global \facebook social graph.

Other works that have measured properties of \facebook include~\cite{Bala08Privacy, Bonneau2009, Viswanath2009, pokingfacebook}.  In~\cite{Bala08Privacy}, Krishnamurthy \etal examine the usage of privacy settings in \myspace and \facebook, and the potential privacy leakage in OSNs. Compared to that work, we have one common privacy attribute, ``View friends``, for which we observe similar results using our unbiased sample. We also have additional privacy settings and the one-hop neighborhood for every node, which allows us to analyze user properties conditioned on their privacy awareness. Bonneau \etal~\cite{Bonneau2009} demonstrate that many interesting user properties can be accurately approximated just by crawling ``public search listings''.

Finally, there is a large body of work on the collection and analysis of datasets for platforms or services that are not pure online social networks but include social networking features. To mention a few examples, Liben-Nowell \etal~\cite{Liben-Nowell-PNAS-05}  studied the \livejournal online community and showed a strong relationship between friendship and geography in social networks.  Cha \etal~\cite{Cha-IMC-07} presented a data-driven analysis of user generated content video popularity distributions by using data collected from \youtube and \daum. Gill \etal~\cite{youtube} also studied a wide range of features of \youtube traffic, including usage patterns, file properties, popularity and referencing characteristics. In~\cite{Gjoka2011_multigraph_JSAC}, we  crawl {\tt Last.FM} a music site with social networking features.

\subsection{Our prior and related work.} 
The conference version of this work appeared in~\cite{Gjoka2010}. This paper is revised and extended to include the following materials. 
(i) A detailed discussion of the uniform userID rejection sampling, which is used as ground truth in this work; see Section \ref{sec:fb_uni} and Appendix A.
(ii) An empirical validation of the assumption that the social graph is static in the time scales of the crawl; see Appendix B.
(iii)  A detailed comparison of  MHRW and RWRW methods and the finding that RWRW is more efficient for all practical purposes; see Section \ref{sec:fb_convergence} for an experimental comparison on \facebook and Appendix C for a  comparison  via analysis and simulation. 
(iv) An extended section on the characterization of \facebook based on a representative sample; see Section \ref{sec:fb_characterization} for additional graphs on node properties and topological characteristics, and new results on privacy settings. 
(iv) A comprehensive review of related work in this section. %


%
%
%
%
%
%
%
%

\section{Sampling Methodology}
\label{sec:samplingmethodology}

We consider OSNs, whose social graph can be modeled as a graph $G=(V,E)$, where $V$ is a set of nodes (users) and $E$ is a set of edges.

\subsection{Assumptions} \label{sec:limitations}
We make the following assumptions and discuss the extent to which they hold:

\begin{itemize}

\item[A1] {\em $G$ is undirected.}  This is true in \facebook (its friendship relations are mutual), but in Twitter the edges are directed, which significantly changes the problem~\cite{Henzinger2000,Twitter08,Ribeiro2010a}. 

\item[A2] {\em We are interested only in the publicly available part of~$G$.} 
This is not a big limitation in \facebook, because all the information we collect is publicly available under default privacy settings. 

\item[A3] {\em $G$ is well connected, and/or we can ignore isolated nodes.}
This holds relatively well in \facebook thanks to its high connection density. In contrast, in \lastfm the friendship graph is highly fragmented, 
which may require more sophisticated crawling approaches~\cite{Gjoka2011_multigraph_JSAC}. 

\item[A4] {\em $G$ remains static during the duration of our crawl.}
We argue in Appendix~B that this assumption holds well in \facebook. %

\item[A5] {\em The OSN supports crawling.} This means that on sampling a node~$v$ we learn the identities of all its neighbors. 
It is typically true in OSNs, \eg through some mechanism such as an API call or HTML scraping (both available in \facebook). 
\end{itemize}

\subsection{Goal and applications} 
Our goal is to obtain a uniform  sample (or more generally a probability sample) of OSN users by crawling the social graph.  This is interesting in its own right, as it allows  to estimate frequencies of user attributes such as age, privacy settings etc.
Furthermore, a probability sample of users allows us to estimate some local topological properties such as node degree distribution, clustering and assortativity. In Section \ref{sec:fb_characterization_topological}, we compute the last two properties based on the one-hop neighborhood of nodes. 
Therefore, a random sample of nodes, obtained using our methodology, is a useful building block towards characterizing structural properties.

We would like to emphasize, however, that a sample of nodes cannot be directly used to obtain a ``representative topology'' for estimating global structural properties.
For example, the nodes and edges in the sample, possibly together with their neighbors (nodes and edges in the egonets) do {\em not} necessarily provide a graph representative of  the entire \facebook with respect to properties such as \eg geodesics.  Therefore, if global structural properties rather than local properties or user attributes are of interest, our node sampling needs to be combined with other techniques 
such as matrix completion  \cite{haupt2008compressed} or block modeling \cite{kurant11_coarsetopology}.

\subsection{Sampling via crawling}

The process of crawling the social graph starts with an initially selected node and proceeds iteratively. In every operation, we visit a node and discover all its neighbors. There are many ways in which we can proceed, depending on which neighbor we choose to visit next. In this section, we describe the sampling methods implemented and compared in this paper. 

\subsubsection{Breadth First Search (BFS)}
At each new iteration the earliest explored but not-yet-visited node is selected next. As this method discovers all nodes within some distance from the starting point, an incomplete BFS is likely to densely cover only some specific region of the graph.

\subsubsection{Random Walk (RW)}
In the classic random walk \cite{Lovasz93}, the next-hop node~$w$ is chosen uniformly at random among the neighbors of the current node~$v$. {\em I.e.,} the probability of moving from $v$ to $w$ is
\begin{displaymath}
P^{\scriptscriptstyle RW}_{v,w} = \left\{ \begin{array}{ll}
\frac{1}{k_v} & \textrm{if $w$ is a neighbor of $v$,} \\
0 & \textrm{otherwise}.
\end{array} \right.
\end{displaymath}
RW is inherently biased. %
Assuming a connected graph and aperiodicity, the probability of being at the particular node $v$ converges to the stationary distribution
$\pi^{\scriptscriptstyle RW}_v = \frac{k_v}{2\cdot|E|}$, {\em i.e.} the classic RW samples nodes w.p.
$\pi^{\scriptscriptstyle RW}_v \sim k_v$.
This is clearly biased towards high degree nodes; {\em e.g.,} a node with twice the degree will be visited by RW twice more often. In Section \ref{sec:fb_evalmethod}, we show that several other node properties
 are correlated with the node degree and thus estimated with bias by RW sampling.

\subsubsection{Re-Weighted Random Walk (RWRW)}\label{subsec:Re-Weighted Random Walk (RWRW)}

A natural next step is to crawl the network using RW, but to correct for the degree bias by an appropriate re-weighting of the measured values. This can be done using the Hansen-Hurwitz 
 estimator \footnote{The simple estimators we use in this paper, e.g., see  Eq. (1), are Hansen-Hurwitz estimators,  which are well-known  to have good properties (consistent and unbiased) under mild conditions; see \cite{kurant11_coarsetopology} for proof of consistency.} \cite{HansenHurwitz1943} as first shown in \cite{Salganik2004, VolzHeckathorn08} for random walks
and also later used in \cite{Rasti09-RDS}. Consider a stationary random walk that has visited $V={v_1,...v_n}$ unique nodes. Each node can belong to one of $m$ groups with respect to a property of interest $A$, which might be the degree, network size or any other discrete-valued node property. Let $(A_1, A_2, .., A_m)$ be all possible values of $A$ and corresponding groups; $\cup_{1}^{m} A_i = V$. {\em E.g.,} if the property of interest is the node degree, $A_i$ contains all nodes $u$ that have degree $k_u=i$. %
 To estimate the probability distribution of $A$, we need to estimate the proportion of nodes with value $A_i,~i=1,..m$:
\begin{equation}\label{hh_equation}
\hat{p}(A_i) = \frac{\sum_{u \in A_i}{1/k_u} }{ \sum_{u \in V}{1/k_u} }
\end{equation}
Estimators for continuous properties can be obtained using related methods, \eg kernel density estimators.

\subsubsection{Metropolis-Hastings Random Walk (MHRW)} \label{subsec:Metropolis-Hastings Random Walk (MHRW)}
Instead of correcting the bias after the walk, one can appropriately modify the transition probabilities so that the walk converges to the desired uniform distribution.
The Metropolis-Hastings algorithm \cite{Metropolis1953} is a general Markov Chain Monte Carlo (MCMC) technique \cite{mcmc-book}  for sampling from a probability distribution $\mu$ that is difficult to sample from directly. In our case, we would like to sample nodes from the uniform distribution $\mu_v = \frac{1}{|V|}$. This can be achieved by the following transition probability:
\begin{displaymath}
P^{\scriptscriptstyle M\!H}_{v,w} = \left\{ \begin{array}{ll}
\min(\frac{1}{k_v}, \frac{1}{k_w}) & \textrm{if $w$ is a neighbor of $v$,} \\
1- \sum_{y\neq v} P^{\scriptscriptstyle M\!H}_{v,y} & \textrm{if $w=v$,} \\
0 & \textrm{otherwise}.
\end{array} \right.
\end{displaymath}
It can be shown that the resulting stationary distribution is $\pi^{\scriptscriptstyle MH}_v = \frac{1}{|V|}$, which is exactly the uniform distribution we are looking for. %
$P^{\scriptscriptstyle M\!H}_{v,w}$ implies the following algorithm, which we refer to simply as MHRW in the rest of the paper: %
\begin{algorithmic}[H]
\small
\STATE {$v \leftarrow $ initial node}.
\WHILE  {stopping criterion not met}
\STATE {Select node $w$ uniformly at random from neighbors~of~$v$.}
\STATE {Generate uniformly at random a number $0\!\leq\!p\!\leq\!1$. }
\IF {  $p \leq \frac{k_v}{k_w}$ }  \STATE { $v\leftarrow w$. }
\ELSE  \STATE { Stay at $v$}
\ENDIF
\ENDWHILE
\end{algorithmic}
At every iteration of MHRW, at the current node $v$ we randomly select a neighbor $w$ and move there w.p. $\min(1,\frac{k_v}{k_w})$. We always accept the move towards a node of smaller degree, and reject some of the moves towards higher degree nodes. This eliminates the bias towards high degree nodes.

\subsection{\label{sec:fb_uni}Ground Truth: Uniform Sample of UserIDs (UNI)}
Assessing the quality of any graph  sampling method on an unknown graph, as it is the case when measuring real systems, is a challenging task. In order to have a ``ground truth'' to compare against, the performance of such methods is typically tested on artificial graphs. 

Fortunately, \facebook  was an exception during the time period we performed our measurements. We capitalized on  a unique opportunity to obtain a uniform sample of \facebook users by generating uniformly random 32-bit userIDs, and by polling \facebook about their existence. If the userID exists (\ie belongs to a valid user), we keep it, otherwise we discard it.
This simple method is a textbook technique known as {\em rejection sampling} \cite{leon-garcia} and in general it allows to sample from any distribution of interest, which in our case is the uniform. In particular, it guarantees to select uniformly random userIDs from the allocated \facebook users regardless of their actual distribution in the userID space,  even when the userIDs are not allocated sequentially or evenly across the userID space. 
For completeness, we re-derive this property in Appendix \ref{sec:appendix_UNIsampling}.  We refer to this method as ``UNI'' and use it as a ground-truth uniform sampler.

Although UNI sampling solves the problem of uniform node sampling in \facebook, crawling remains important. 
Indeed, the userID space must not be sparsely allocated for UNI to be efficient. During our data collection (April-May 2009) the number of \facebook users  ($\sim 200\times10^{6}$)  was comparable to the size of the userID space ($2^{32} \sim 4.3\times10^{9}$), resulting in about one user retrieved per 22 attempts on average. If the userID were 64 bit long or consisting of strings of arbitrary length, UNI would had been infeasible.\footnote{To mention a few such cases in the same time frame: \orkut had  a 64bit userID and hi5 used a concatenation of userID+Name. Interestingly, within days to weeks after our measurements were completed, \facebook changed its userID allocation space from  32 bit to 64 bit \cite{userid64bit}. 
Section \ref{subsection:userID} contains more information about userID space usage in \facebook in April 2009.}

In summary, we were fortunate to be able to obtain a uniform independence sample of userIDs, which we then used as a baseline for comparison (our ``ground truth'') and showed that our results conform closely to it. However, crawling friendship relations is a fundamental primitive available in all OSNs and, we believe, the right building block for designing sampling techniques in OSNs in the general case. 

\subsection{Convergence}

\subsubsection{\label{sec:fb_multiple-chains}Using Multiple Parallel Walks}

Multiple parallel walks are used in the MCMC literature \cite{mcmc-book} to improve convergence. Intuitively, if we only have one walk, the walk may get trapped in  cluster while exploring the graph, which may lead to erroneous diagnosis of convergence. Having multiple parallel walks reduces the  probability of this happening and allows for more accurate convergence diagnostics. 
An additional advantage of multiple parallel walks, from an implementation point of view, is that it is amenable to parallel implementation from different machines or different threads in the same machine. 

\subsubsection{Detecting Convergence with Online Diagnostics}
\label{sec:fb_diagnostics}

Valid inferences from MCMC are based on the assumption that the samples are derived from the equilibrium distribution, which is true asymptotically.
In order to correctly diagnose  when convergence to equilibrium occurs, we use standard diagnostic tests developed within the MCMC literature \cite{mcmc-book}.
In particular, we would like to use diagnostic tests to answer at least the following questions:
\begin{itemize}
\item How many of the initial samples in each walk do we need to discard to lose dependence from the starting point (or burn-in) ?
\item How many samples do we need before we have collected a representative sample?
\end{itemize}

A standard approach is to run the sampling long enough and to discard a number of initial burn-in samples proactively. From a practical point of view, however, the burn-in comes at a cost. In the case of \facebook, it is the consumed bandwidth (in the order of gigabytes) and measurement time (days or weeks). It is therefore crucial to assess the convergence of our MCMC sampling, and to decide on appropriate settings of burn-in and total running time.

Given that during a crawl we do not know the target distribution, we can only estimate convergence from the statistical properties of the walks as they are collected. Here we present two standard convergence tests, widely accepted and well documented in the MCMC literature, Geweke \cite{geweke} and Gelman-Rubin \cite{gelman-rubin}, described below.  
 In Section \ref{sec:fb_evalmethod}, we apply these tests on several node properties, including the node degree, userID, network ID and membership in a specific network; please see Section \ref{sec:fb_metrics} for details.
Below, we briefly outline the rationale of these tests. %

\textbf{{\em Geweke Diagnostic.}} The Geweke diagnostic \cite{geweke} detects the convergence of a single Markov chain. Let $X$ be a single sequence of samples of our metric of interest. Geweke considers two subsequences of $X$, its beginning $X_a$ (typically the first 10\%), and its end $X_b$ (typically the last 50\%). Based on $X_a$ and $X_b$, we compute the z-statistic:
$$ z = \frac{E(X_a) - E(X_b) }{\sqrt{ Var(X_a) + Var(X_b) } }$$
With increasing number of iterations,  $X_a$ and $X_b$ move further apart, which limits the correlation between them. As they measure the same metric, they should be identically distributed when converged and, according to the law of large numbers, the $z$ values become normally distributed with mean 0 and variance 1. We can declare convergence when all values fall in the $[-1,1]$ interval.

\textbf{{\em Gelman-Rubin Diagnostic.}} Monitoring one long sequence of nodes has some disadvantages. For example, if our chain stays long enough in some non-representative region of the parameter space, we might erroneously declare convergence. For this reason, Gelman and Rubin~\cite{gelman-rubin} proposed to monitor $m>1$ sequences. Intuitively speaking, the Gelman-Rubin diagnostic compares the empirical distributions of individual chains with the empirical distribution of all sequences together: if these two are similar, we declare convergence. The test outputs a single value $R$ that is a function of means and variances of all chains. With time, $R$ approaches 1, and convergence is declared typically for values smaller than 1.02.

\section{Data Collection}
\label{sec:fb_datacollection}

\begin{figure}[t]
\centering
\includegraphics[width=0.45\textwidth]{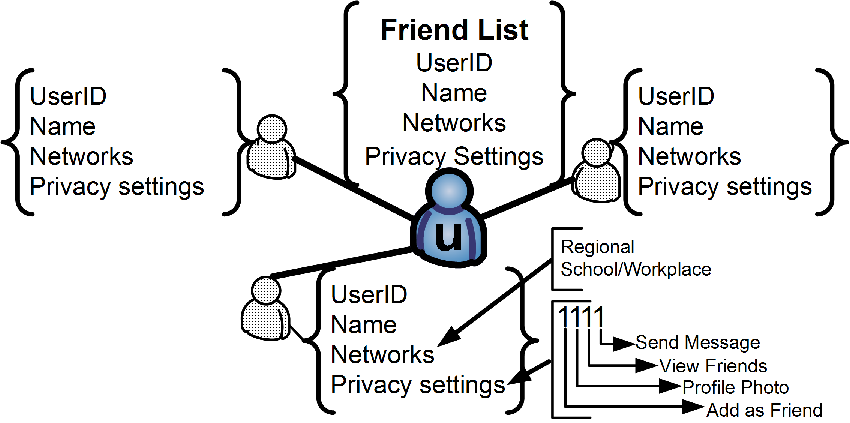}
\caption{Basic node information collected when visiting a user $u$.}
\label{fig:nodeinfo}
\end{figure}

\begin{table}
\centering
\begin{tabular}{@{}l|l|l}
  bit  & attribute & explanation\\
 \hline
  1 & Add as friend & =1 if $w$ can propose to `friend' $u$ \\
  2 & Photo &  =1 if $w$ can see the profile photo of $u$ \\
  3 & View friends & =1 if $w$ can see the friends of $u$ \\
  4 & Send message & =1 if $w$ can send a message to $u$ \\
\end{tabular}
\caption{Privacy settings of a user $u$ with respect to her non-friend $w$.}
\label{tab:Privacy settings}
\end{table}

\subsection{User properties of interest}

Fig.~\ref{fig:nodeinfo} summarizes the information collected when visiting the ``show friends'' web page of a sampled user $u$, which we refer to as {\em basic node information}.

{\bf Name and userID.}
Each user is uniquely defined by her userID, which is a 32-bit  number\footnote{\facebook changed to 64-bit user ID space after May 2009 \cite{userid64bit} whereas our crawls were collected during April-May 2009.}. Each user presumably provides her  real name. The names do not have to be unique.

{\bf Friends list.}
A core idea in social networks is the possibility to declare friendship between users. In \facebook, friendship is always mutual and must be accepted by both sides. Thus the social network is undirected.

{\bf Networks.}
\label{subsec:fb_networks}
\facebook uses two types of ``networks'' to organize its users. 
The first are  \emph{regional} (geographical) networks\footnote{Regional networks were available at the time of this study but were phased out starting from June 2009\cite{fb-regional}}. There are 507 predefined regional networks that correspond to cities, regions, and countries around the world. A user can freely join any regional network but can be a member of only one regional network at a time. Changes are allowed, but no more than twice every 6 months (April 2009).
The second type of networks contain user affiliations with colleges, workplaces, and high schools and have stricter membership criteria: they require a valid email account from the corresponding domain, \eg to join the UC Irvine network you have to provide a ``@uci.edu'' email account. 
A user can belong to many networks of the second type.

{\bf Privacy settings $Q_v$.} Each user $u$ can restrict the amount of information revealed to any non-friend node $w$, as well as the possibility of interaction with $w$.
These are captured by four basic binary privacy attributes, as described in Table~\ref{tab:Privacy settings}.
We refer to the resulting 4-bit number as privacy settings $Q_v$ of node $v$. By default, \facebook sets $Q_v=1111$ (allow all).

{\bf Friends of $u$.} The ``show friends'' web page of user $u$ exposes network membership information and privacy settings for each listed friend. Therefore, we collect such information for all friends of $u$, at no additional cost.

{\bf Profiles.}
Much more information about a user can potentially be obtained by viewing her profile. Unless restricted by the user, the profile can be displayed by her friends and users from the same network.
In this work, we do not collect any profile information, even if it is publicly available. We study only the basic node information shown in Fig.\ref{fig:nodeinfo}.

\begin{figure}[t]
\centering
\includegraphics[width=0.45\textwidth]{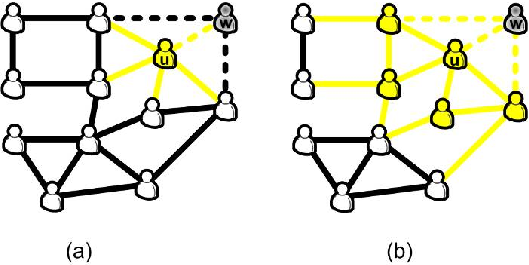}
\caption{ (a) Sampled user $u$ with observed edges in yellow color. (b) The extended ego network of user $u$ with observed nodes and edges in yellow color. Invalid neighbor $w$, whose privacy settings $Q_w=**0*$ do not allow friend listing, is discarded.}
\label{fig:egonet}
\end{figure}

{\bf Ego Networks.}
The sample of nodes collected by our method enables us to study many features of FB users in a statistically unbiased manner. However,  more elaborate topological measures, such as clustering coefficient and assortativity, cannot be easily estimated based purely on a single-node view. For this reason, we decided to also collect a number of \emph{extended ego nets} \footnote{ We use the extended egonet sample differently from the Random Node Neighbor (RNN) sample presented in \cite{Leskovec2006_sampling_from_large_graphs}. We are interested in estimating properties of the ego nodes only whereas RNN \cite{Leskovec2006_sampling_from_large_graphs} looks at the induced subgraph of all sampled nodes and is interested in estimating properties of the ego and all alters. Therefore, the sample of extended egonets, that we collect in this work, is expected to capture very well community structure properties (\ie clustering coefficient) of the whole Facebook population. }  (see Fig~\ref{fig:egonet}), for $\sim$37K ``ego'' nodes, randomly selected from all nodes in MHRW. 

\begin{table*}[th!]
\footnotesize
\begin{minipage}{4in}
\begin{tabular}{@{}l|@{}l|@{}l|@{}l|@{}l}
Crawling method                                             & MHRW            &   RW          &  BFS          & UNI\\
\hline
Total number of valid users                   & 28$\times$81K   & 28$\times$81K & 28$\times$81K  &  984K\\
Total number of \textit{unique} users         & 957K            & 2.19M         &  2.20M        &  984K\\
Total number of \textit{unique} neighbors     & 72.2M           & 120.1M         & 96M        &  58.4M\\
Crawling period                               &  04/18-04/23    & 05/03-05/08   & 04/30-05/03   &  04/22-04/30 \\
\hline
Avg Degree                                    & 95.2            & 338            & 323       & 94.1 \\
Median Degree                                 & 40              & 234            & 208         & 38 \\
\end{tabular}
\end{minipage}
\begin{minipage}{3in}
\footnotesize
 \begin{center}

\begin{tabular}{@{}l|@{}l}
& Number of overlapping users\\
\hline
MHRW $\cap$ RW       & 16.2K\\
\hline
MHRW $\cap$ BFS      & 15.1K\\
\hline
MHRW $\cap$ Uniform  & 4.1K\\
\hline
RW $\cap$ BFS        & 64.2K\\
\hline
RW $\cap$ Uniform    & 9.3K\\
\hline
BFS $\cap$ Uniform   & 15.1K
\end{tabular}
\end{center}
\end{minipage}

\caption{(Left:) Datasets collected by MHRW, RW, BFS and UNI in 2009. (Right:)The overlap between different datasets is small.
}

\label{tab:datasets}
\end{table*}

\subsection{Crawling Process}

\begin{figure}
\centering
\includegraphics[width=0.35\textwidth]{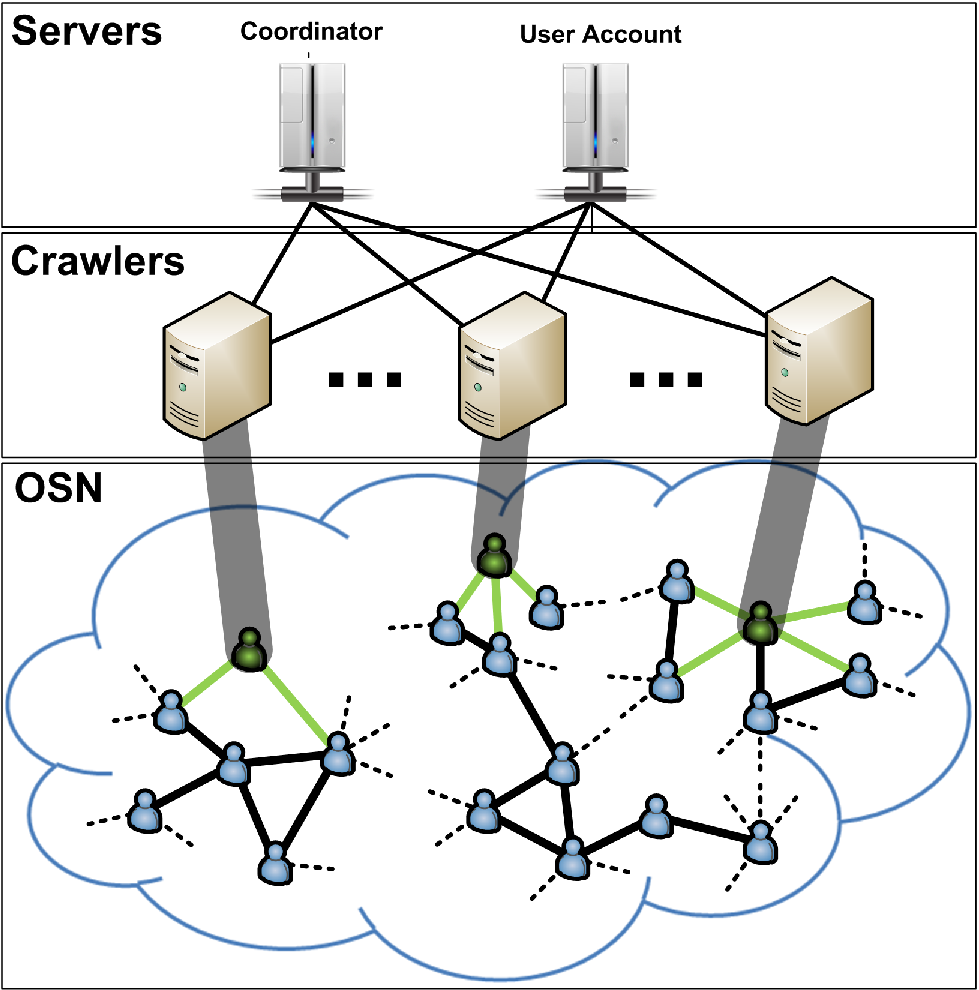}
\caption{Distributed Crawling of an Online Social Network}
\label{fig:fb_crawlingdiagram}
\end{figure}

In order to apply our methodology to real-life  OSNs, we implemented high-performance distributed crawlers  that explored the social graph in a systematic and efficient way.

\subsubsection{Challenges}
There are several practical challenges we faced while crawling the social graph of OSNs.
First, OSNs usually use some defense mechanisms against automated data mining. 
Mislove \etal \cite{Mislove2007} reported rate limits per IP while crawling \orkut. Similarly, in our \facebook crawls we experienced banned accounts, probably due to excessive traffic. 
Second, in \facebook, 
the API calls are usually more restrictive than HTML scraping,  %
which forced us to implement the latter.
Third, many modern web sites enable asynchronous loading of web content (\ie use AJAX), which requires more sophisticated customization of the crawlers. 
Finally, in order to satisfy assumption A4, the data collection time should be relatively small, in the order of a few days 
(see Appendix~B).

\subsubsection{Implementation}

Fig~\ref{fig:fb_crawlingdiagram} depicts an overview of our distributed crawling process.

First, we use a large number of machines with limited memory (100 Mbytes-1GBytes RAM) and disk space (up to 5GBytes), to parallelize our crawling and shorten the data collection time. We have up to three layers of parallelism in each machine. Each crawling machine runs one or more crawling processes. Each crawling process shares one user account between multiple crawling threads within it. Each crawling thread fetches data asynchronously where possible.  

Second, we use one machine as coordinator server that 
(i)~controls the number of connections or amount of bandwidth over the whole cluster, %
(ii)~keeps track of already fetched users to avoid fetching duplicate data, and
(iii)~maintains a data structure that stores the crawl frontier, \eg a queue for BFS. 
%

Third, a user account server stores the login/API accounts, created manually by the administrator. 
When a crawling process is initiated, it requests an unused account from the user account server; the crawling process is activated only if a valid account is available.  

\subsubsection{Invalid users}
There are two types of users that we declare as \textit{invalid}. First, if a user $u$ decides to hide her friends and to set the privacy settings to $Q_u=**0*$, the crawl cannot continue. We address this problem by backtracking to the previous node and continuing the crawl from there, as if $u$ was never selected.
Second, there exist nodes with degree $k_v=0$; these are not reachable by any crawls, but we stumble upon them during the UNI sampling of the userID space.
Discarding both types of nodes is consistent with our assumptions (A2, A3), where we already declared that we exclude such nodes (either not publicly available (A2) or isolated (A3)) from the graph we want to sample.

\subsubsection{Execution of crawls}

We ran $28$ different independent crawls for each crawling methodology, namely MHRW, BFS and RW, all seeded at the same set of randomly selected nodes. 
We collected exactly 81K samples for each independent crawl. 
We count towards this value all repetitions, such as the self-transitions of MHRW, and returning to an already visited state (RW and MHRW).  
In addition to the 28$\times$3 crawls (BFS, RW and MHRW), we ran the UNI sampling until we collected  984K valid users, which is comparable to the 957K unique users collected with MHRW.

\subsection{Description of Datasets}

Table~\ref{tab:datasets} summarizes the datasets collected using the crawling techniques under comparison. This information refers to all sampled nodes, before discarding any burn-in. 
For each of MHRW, RW, and BFS, we collected the total of $28 \times 81K=2.26M$ nodes. 
However, because MHRW and RW sample with repetitions, and because the 28 BFSes may partially overlap, the number of \emph{unique} nodes sampled by these methods is smaller. This effect is especially visible under MHRW that collected only 957K unique nodes.
Table \ref{tab:datasets}(right) shows that the percentage of common users between the MHRW, RW, BFS and UNI datasets is very small, as expected. The largest observed, but still objectively small, overlap is between RW and BFS and is probably due to the common starting points selected.

To collect the UNI dataset, we checked $\sim$ 18.5M user IDs picked uniformly at random from $[1, 2^{32}]$.  Out of them, only 1,216K users existed. Among them, 228K users had zero friends; we discarded these isolated users to be consistent with our problem statement. This results in a set of 984K valid users with at least one friend each. 

To analyze topological characteristics of the \facebook population, we collected $\sim$ 37K egonets that contain basic node information (see Fig \ref{fig:nodeinfo}) for  $\sim$ 5.8M unique neighbors. 
Table~\ref{tab:egonet} contains a summary of the egonet dataset, including properties that we analyze in section \ref{sec:fb_characterization}.

\begin{table}[h]
\centering
\begin{tabular}{l|l}
\hline
Number of egonets              &  37K      \\
Number of neighbors            &  9.3M      \\
Number of unique neighbors     &  5.8M      \\
Crawling period                &   04/24-05/01  \\
\hline
Avg Clustering coefficient  &  0.16\\
Avg Assortativity           &  0.233\\
\end{tabular}
\caption{Ego networks collected for 37K nodes,  randomly selected from the users in the MHRW dataset.}
\label{tab:egonet}
\end{table}

Overall, 
we sampled 11.6 million unique nodes, and observed other 160M as their (unsampled) neighbors. 
This is a very large sample by itself, especially given that \facebook had reported having close to 200 million active users during the time of these measurements. %

\section{Evaluation of Sampling Techniques}
\label{sec:fb_evalmethod}

In this section, we evaluate all candidate crawling techniques (namely BFS, RW and RWRW, MHRW), in terms of their efficiency (convergence) and quality
(estimation bias).
In Section \ref{sec:fb_convergence}, we study the convergence of the random walk methods with respect to several properties of interest. We find a burn-in period of 6K samples, which we exclude from each independent crawl. The remaining 75K x 28 sampled nodes is our main sample dataset; for a fair comparison we also exclude the same number of burn-in samples from all datasets. In Section \ref{sec:fb_uniformity} we examine the quality of the estimation based on each sample. In Section \ref{sec:fb_recommendation}, we summarize our findings and provide practical recommendations.

\subsection{\label{sec:fb_convergence}Convergence analysis}

There are several crucial parameters that affect the convergence of a Markov Chain sample. In this section, we study these parameters by (i) applying formal convergence tests and (ii) using simple, yet insightful, visual inspection of the related traces and histograms.

\subsubsection{Burn-in} 
For the random walk-based methods,  a number of samples need to be discarded to lose dependence on the initial seed point. Since there is a cost for every user we sample, we would like to choose this value using formal convergence diagnostics so as not to waste resources. Here, we apply the convergence diagnostics presented in Section \ref{sec:fb_diagnostics} to several properties of the sampled nodes  and choose as burn-in the maximum period from all tests.

\begin{figure}[bht]
\centering
\subfigure[Number of friends]{
\includegraphics[width=0.30\textwidth,angle=270]{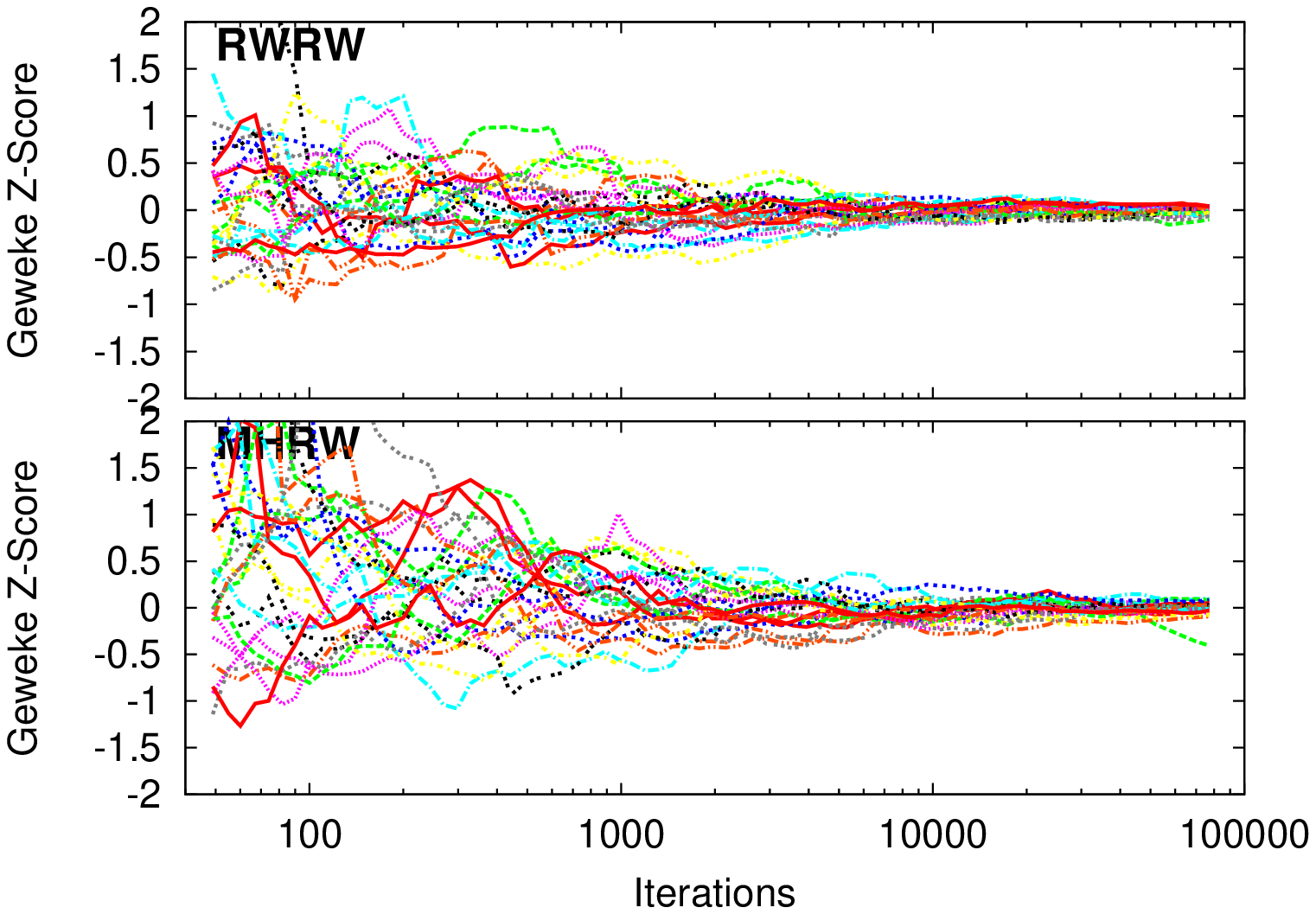}}
\subfigure[Regional network ID]{
\includegraphics[width=0.30\textwidth,angle=270]{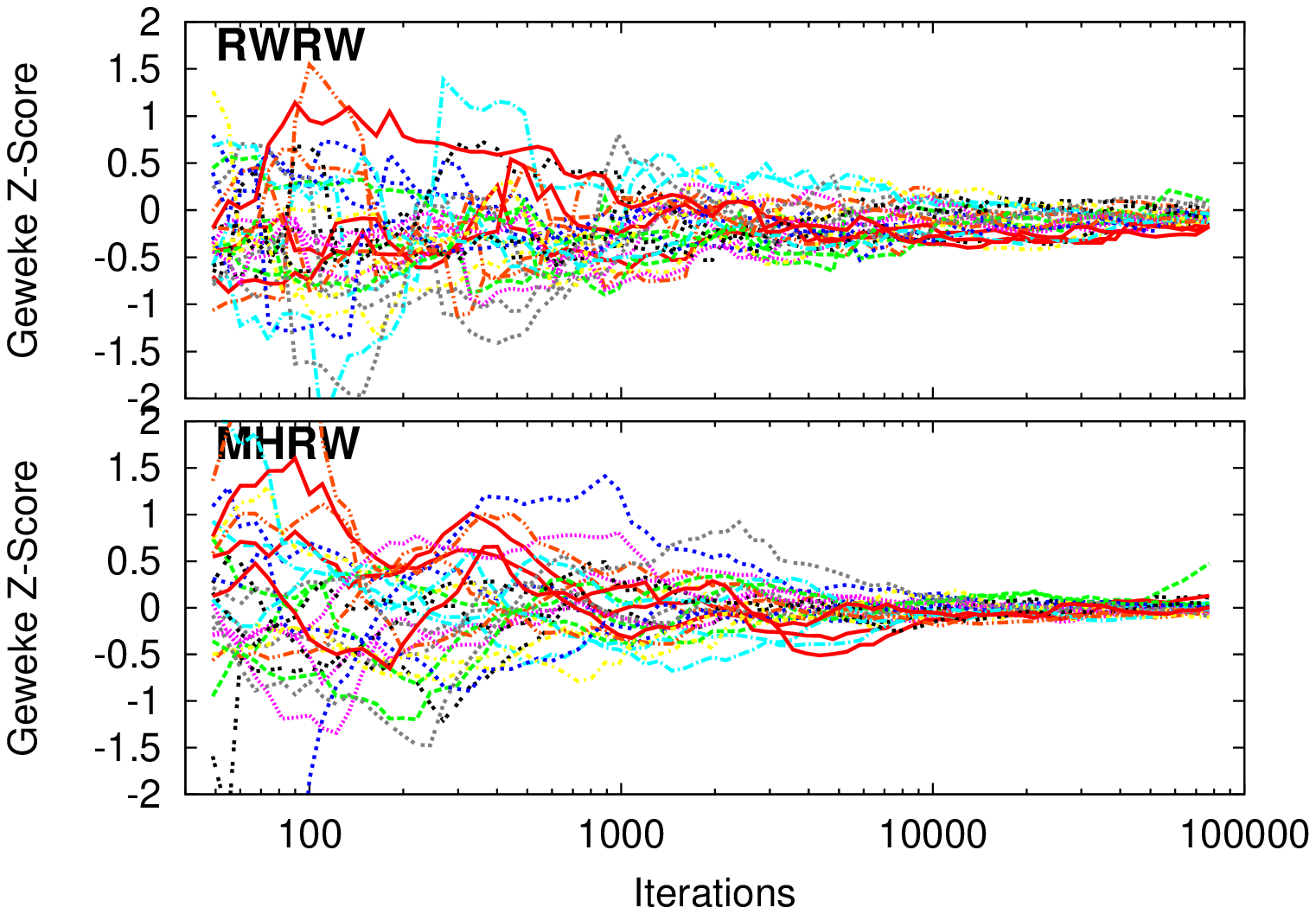}}
\caption{Geweke z score (0K..81K) for number of friends (top) and regional network affiliation (bottom). Each line shows the Geweke score for each of the 28 parallel walks.
}
\label{fig:gewekeresults_burnin}
\end{figure}

The Geweke diagnostic is applied in each of the 28 walks separately and compares the difference between the first 10\%  and the last 50\% of the walk samples. It declares convergence when all 28 values fall in the $[-1,1]$ interval. Fig.~\ref{fig:gewekeresults_burnin} presents the results of the Geweke diagnostic for the user properties of node degree and regional network membership. We start at 50 iterations and plot 150 points logarithmically spaced. We observe that after approximately $500-2000$ iterations we have a z-score strictly between $[-1,1]$. We also see that RWRW and MHRW perform similarly w.r.t. the Geweke diagnostic.

\begin{figure}
\centering
\includegraphics[width=0.30\textwidth,angle=270]{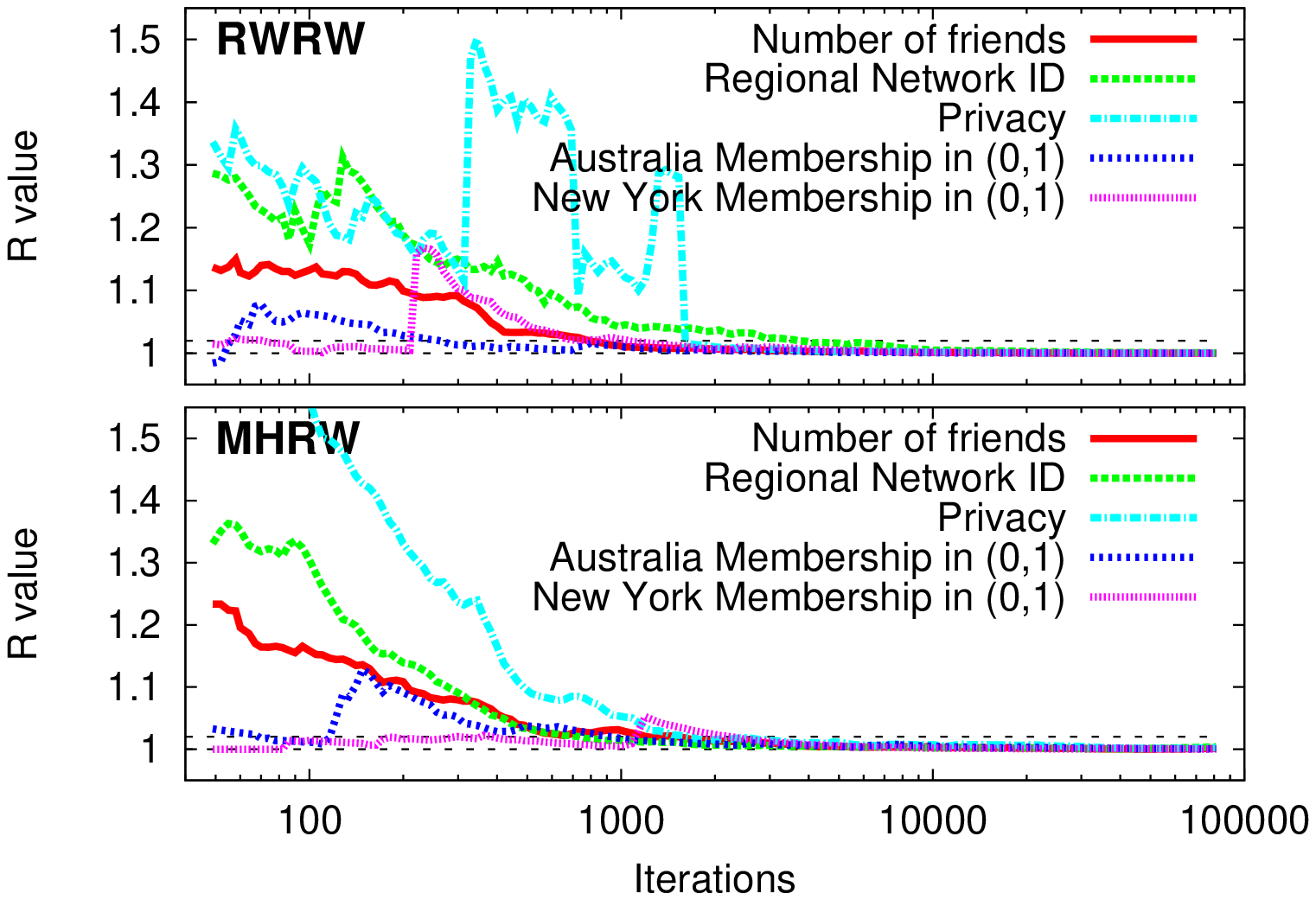}
\caption{ Gelman-Rubin R score (0K..81K) for five different metrics.
}
\label{fig:gelmanrubin_burnin}
\end{figure}

The Gelman-Rubin diagnostic analyzes all the 28 walks at once by summarizing the difference of the between-walk variances and within-walk variances. In Fig.~\ref{fig:gelmanrubin_burnin} we plot the R score for the following metrics (i) number of friends (or node degree) (ii) networkID (or regional network) (iii) privacy settings $Q_v$ (iv) membership in specific regional networks, namely Australia, New York and Colombia. The last user property is defined as follows: if the user in iteration $i$ is a member of network $x$ then the metric is set to $1$, otherwise it is set to $0$. We can see that the R score varies considerably in the initial hundred to thousand iterations for all properties.  To pick an example, we observe a spike between iterations $1,000$ and $2,000$ in the MHRW crawl for the New York membership. This is most likely the result of certain walks getting trapped within the New York network, which is particularly large. Eventually, after $3000$ iterations all the R scores for the properties of interest drop below $1.02$, the typical target value used for convergence indicator.

We declare convergence when all tests have detected it. The Gelman-Rubin test is the last one at 3K nodes. To be even safer, in each independent walk we conservatively discard 6K nodes, out of 81K nodes total. In the remainder of the evaluation, we work only with the remaining 75K nodes per independent chain for RW, RWRW and MHRW.

\subsubsection{Total Running Time} 
Another decision we have to make is about the {\em walk length}, excluding the burn-in samples. This length should be appropriately long to ensure that we are at equilibrium. Here, we utilize multiple ways to analyze the collected samples, so as to increase our confidence that the collected samples are appropriate for further statistical analysis.

\begin{figure}
\centering
\subfigure[Geweke on Number of Friends]{
\includegraphics[width=0.30\textwidth,angle=270]{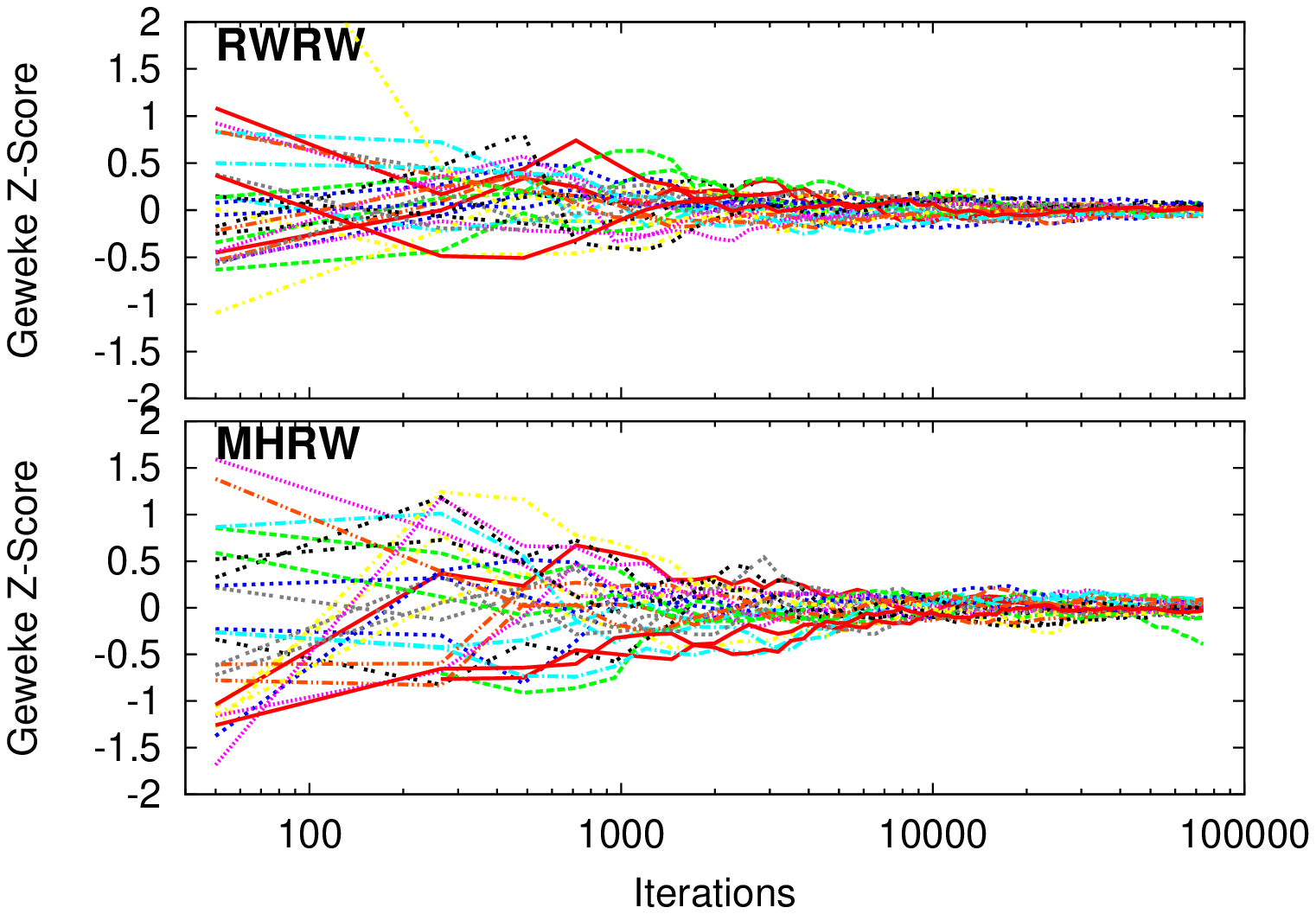}}
\subfigure[Gelman Rubin]{
\includegraphics[width=0.30\textwidth,angle=270]{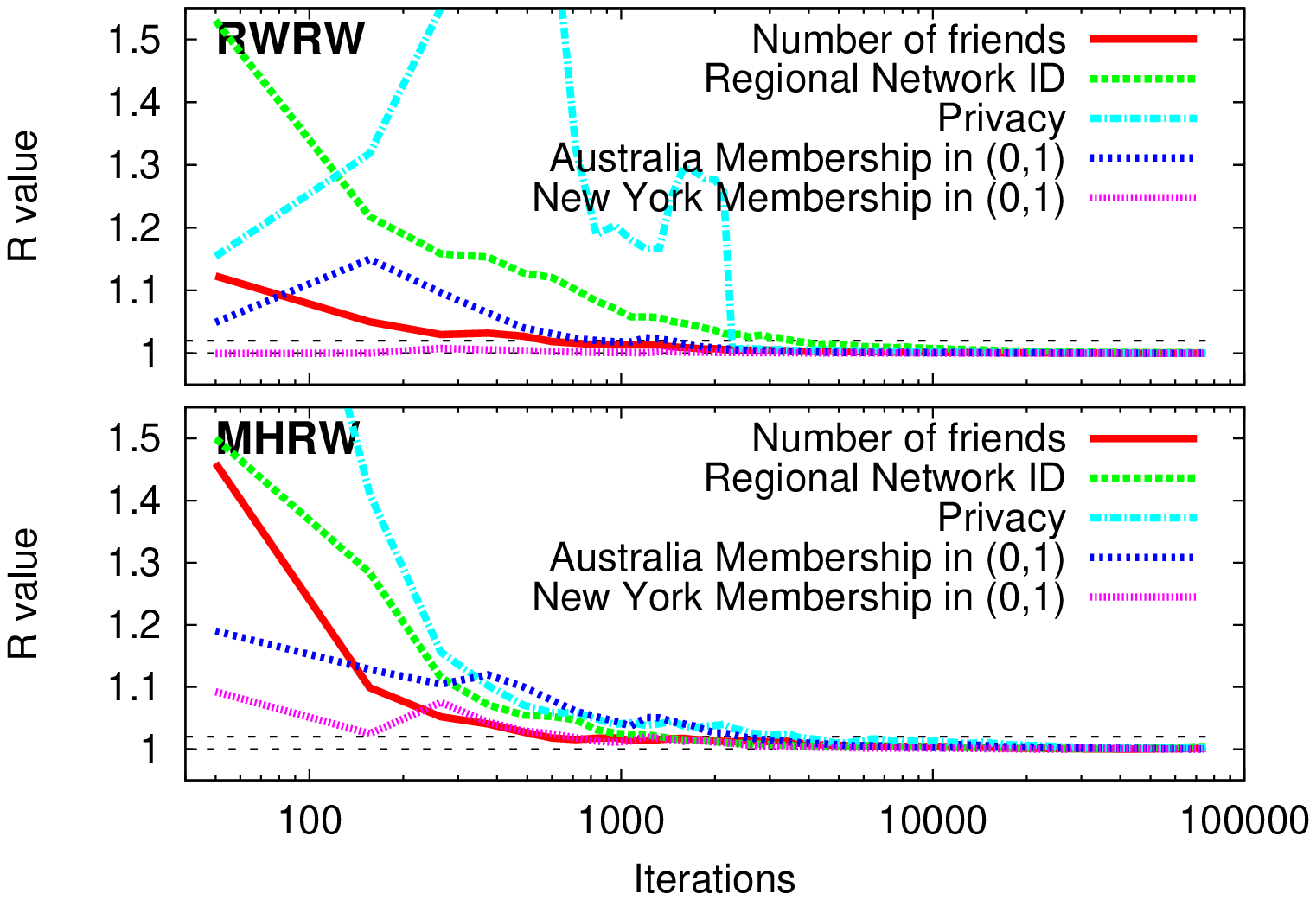}}
\caption{Online convergence diagnostics for samples 6K..81K (without burn-in). (top) Geweke z score for number of friends. (bottom) Gelman-Rubin score for five different metrics.}
\label{fig:convergence_walklength}
\end{figure}

First, we apply formal convergence diagnostics that allow us to assess convergence online by indicating approximate equilibrium. Fig~\ref{fig:convergence_walklength} shows the Geweke z-score for the number of friends (top) and the Gelman-Rubin R score  for five different properties (bottom). These results are obtained after discarding the burn-in samples (0K..6K). They show that convergence is attained with at least 3k samples per walk, similar to the section in which we determined the burn-in. This is an indication that the \facebook social graph is well connected and our random walks achieved good mixing with our initial selection of random seeds.

Second, we perform visual inspection to check the convergence state of our sample by plotting for each walk the running mean of a user property against the iteration number. %
 The intuition  is that if convergence has been reached, the running mean of the property will not drastically change as the number of iterations increases. Fig~\ref{fig:means_concergence} shows for each crawl type the running mean (i)  for the node degree in the UNI sample, (ii) in each of the 28 walks individually, and (iii)  in an average  crawl that combines all 28 walks. It can be seen that in order to estimate the average node degree $\overline{k_v}$ based on only a single MHRW or RW walk, we should take at least 10K iterations to be likely to get within $\pm10\%$ off the real value.  In contrast, averaging over all 28 walks seems to provide similar  or better confidence after fewer than 100 iterations per walk or $100\times28\sim 3k$ samples over all walks. Additionally, the average MHRW and RW crawls reach stability within $350\times 28 \sim 10K$ iterations. It is quite clear that the use of multiple parallel walks is very beneficial in the estimation of user properties of interest.

\begin{figure*}
\centering
\psfrag{E[deg]}[c][t][0.9]{$\overline{k_v}$}
\psfrag{t}[c][b][0.9]{Iterations}
\subfigure[MHRW]{
\includegraphics[width=.23\textwidth]{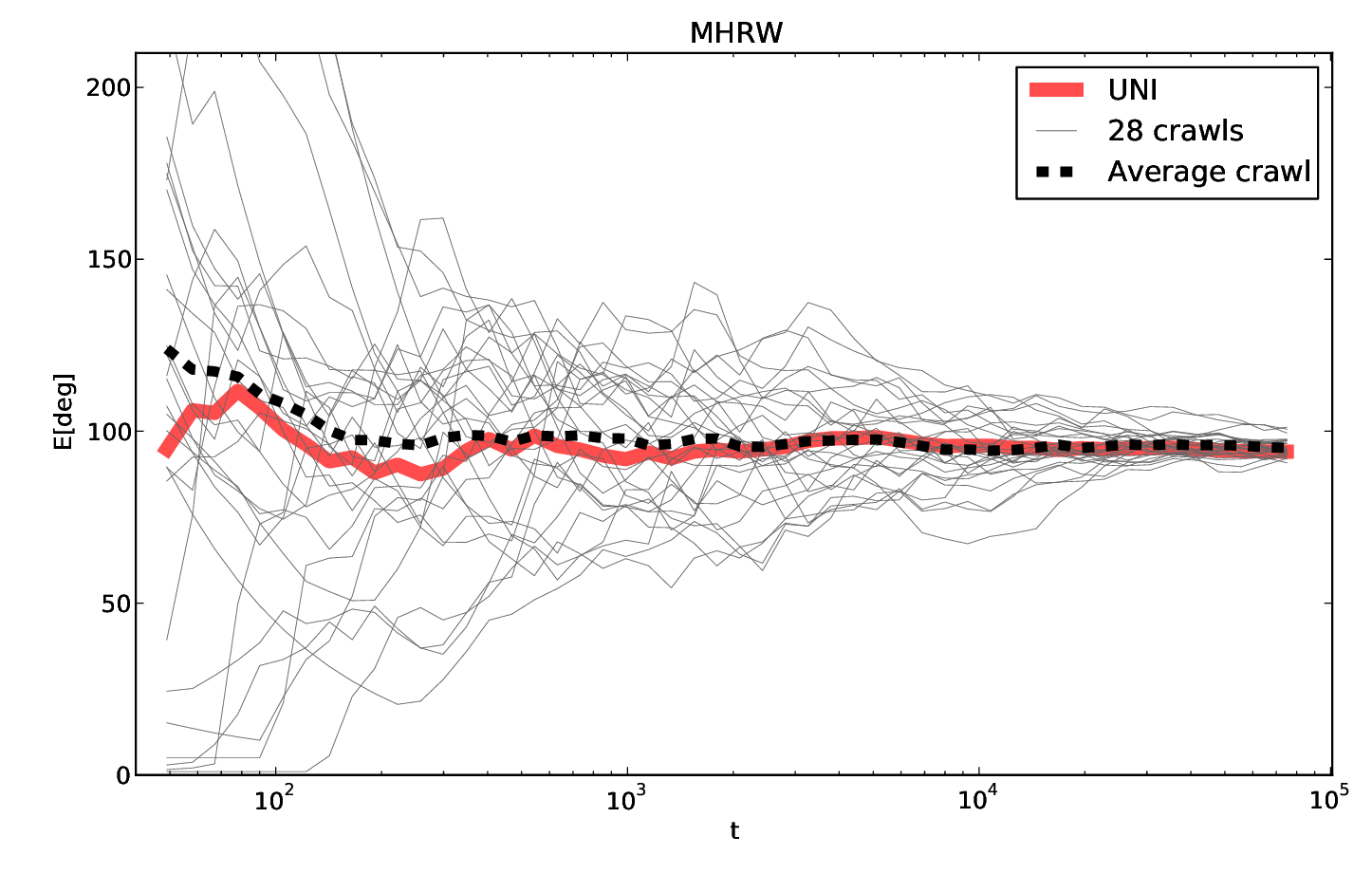}}\hspace{0pt}
\subfigure[RWRW]{
\includegraphics[width=.23\textwidth]{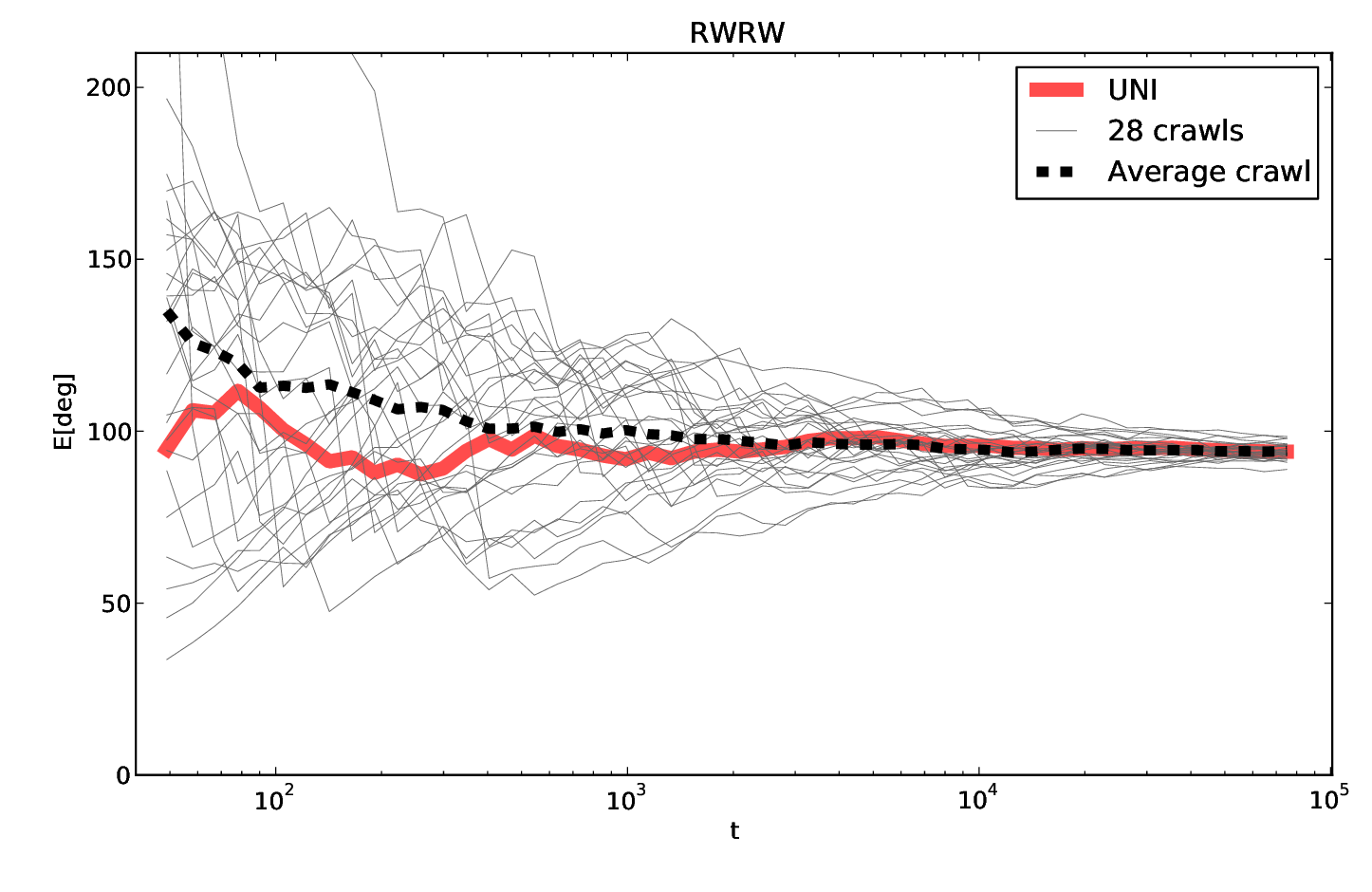}}
\subfigure[RW]{
\includegraphics[width=.23\textwidth]{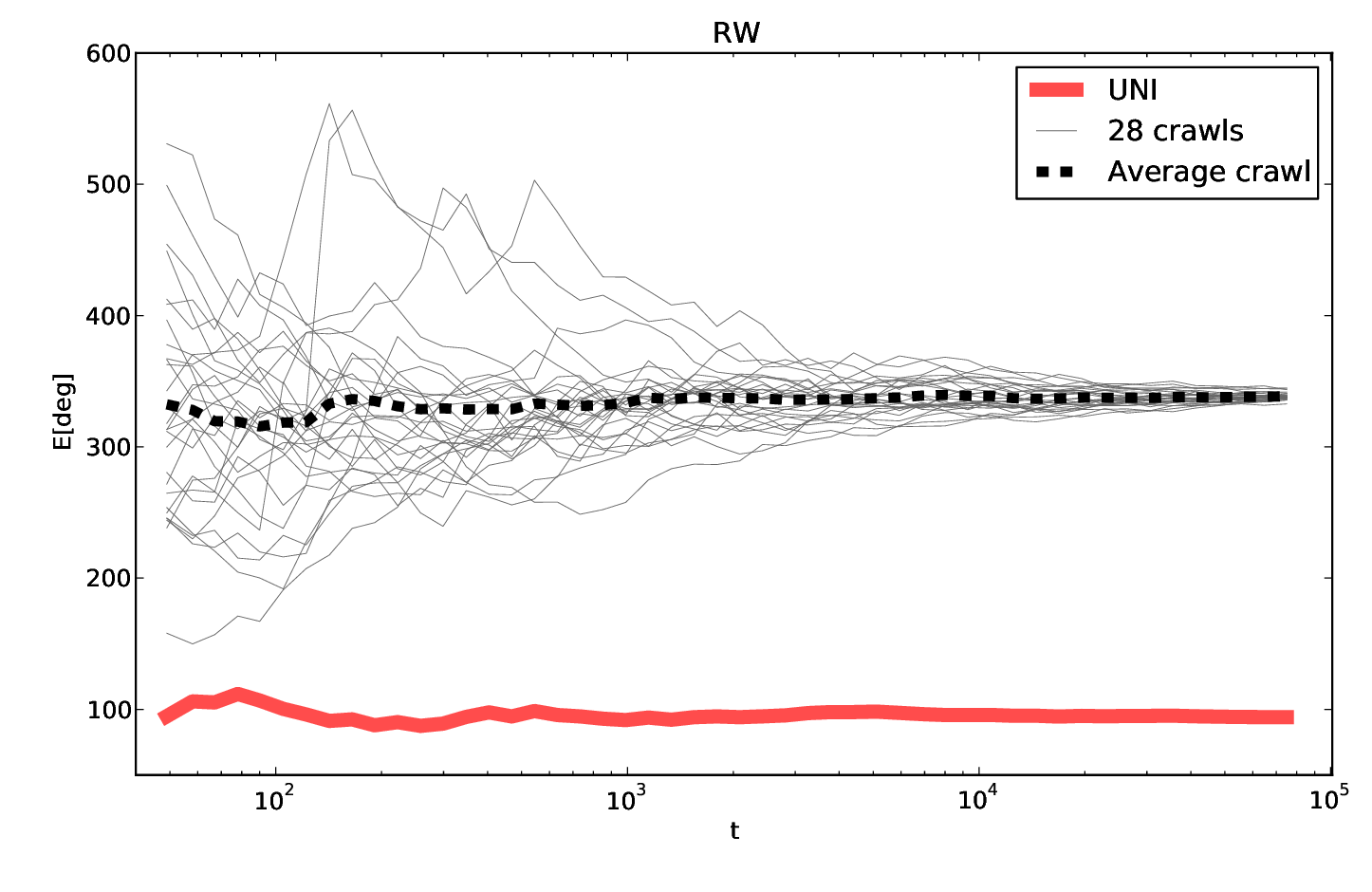}}\hspace{0pt}
\subfigure[BFS]{
\includegraphics[width=.23\textwidth]{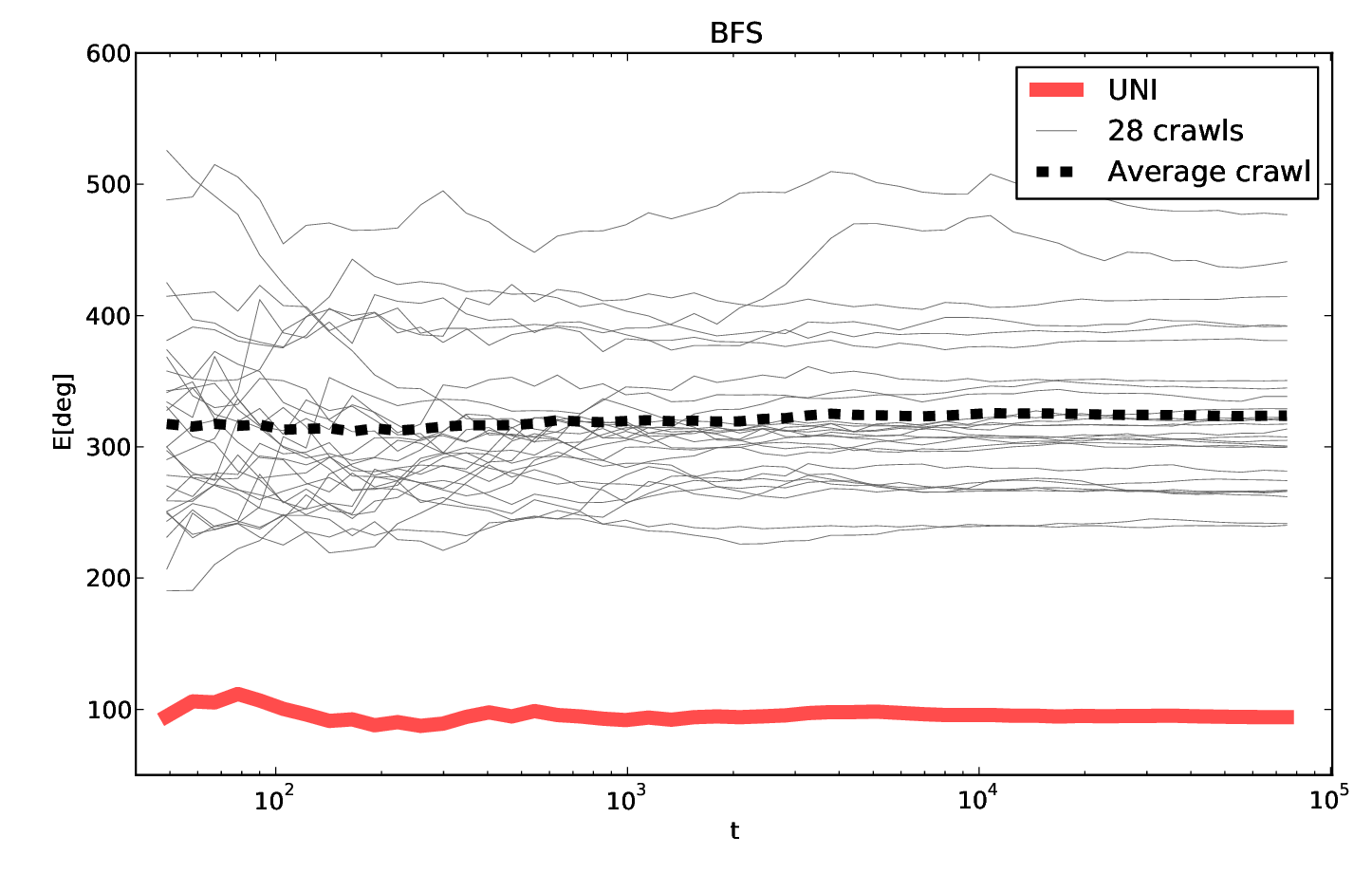}}
\caption{Average node degree $\overline{k_v}$ observed by each crawl, as a function of the number of iterations (or running mean).}
\label{fig:means_concergence}
\end{figure*}

According to the diagnostics  and visual inspection, we need at least 3k samples per walk or $3k\times 28 \sim 84k$ over all walks. Since we were not resource constrained during our crawling, we continued sampling users until we reached 81K per walk. One obvious reason is that more samples should decrease the estimation variance. Another reason is that more samples allow us to break the correlation between consecutive samples by thinning the set of sampled users. We use such a thinning process to collect egonets.

\subsubsection{Thinning} 
Let us examine the effect of a larger sample on the estimation of user properties. Fig.~\ref{fig:histograms_stopTime_thinning.eps} shows the percentage of sampled users with specific node degrees and network affiliations, rather than the average over the entire distribution. A walk length of 75K (top) results in much smaller estimation variance per walk than taking 5K consecutive iterations from 50-55k (middle).
Fig.\ref{fig:histograms_stopTime_thinning.eps} also reveals the correlation between  consecutive samples, even after equilibrium has been reached. It is sometimes reasonable to break this correlation, by considering every $i$th sample, a process which is called \emph{thinning}. %
The bottom plots in Fig.~\ref{fig:histograms_stopTime_thinning.eps} show 5K iterations per walk with a thinning factor of $i=10$. It performs much better than the middle plot, despite the same total number of samples.

\begin{figure}
\centering
\psfrag{Prob(deg)}[c][t][0.9]{$\Prob(k_v=k)$}
\psfrag{node degree}[c][b][0.9]{node degree $k$}
\includegraphics[width=0.47\textwidth]{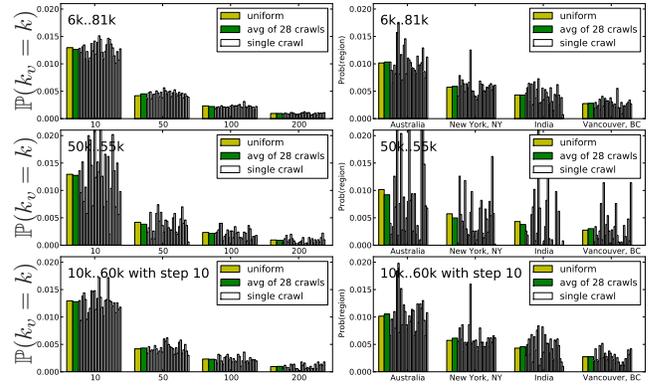}
\caption{The effect of walk length and thinning on the results. We present histograms of visits at nodes with a specific degree $k\in\{10,50,100,200\}$ and network membership (Australia, New York, India, Vancouver), generated under three conditions.\ (top):~All nodes visited after the first 6K burn-in nodes. (middle):~5K consecutive nodes, from hop 50K to hop 55K. This represents a short walk length.\ (bottom):~5k nodes by taking every 10th sample (thinning).
}
\label{fig:histograms_stopTime_thinning.eps}
\end{figure}

Thinning in MCMC samplings has the side advantage of saving space instead of storing all collected samples. In the case of crawling OSNs, the main bottleneck is the time and bandwidth necessary to perform a single transition, rather than storage and post-processing of the extracted information. Therefore we did not apply thinning to our basic crawls.

However, we applied another idea (sub-sampling), that has a similar effect with thinning, when collecting the second part of our data - the egonets. Indeed, in order to collect the information on a single egonet, our crawler had to visit the user and all its friends, an average $\sim100$ nodes. Due to bandwidth and time constraints, we could fetch only 37K egonets. In order to avoid correlations between consecutive egonets, we collected a random sub-sample of the MHRW (post burn-in) sample, which essentially introduced spacing among sub-sampled nodes.

\subsubsection{Comparison to Ground Truth} 
Finally,  we compare the random walk techniques in terms of their distance from the true uniform (UNI) distribution as a function of the iterations. In Fig.~\ref{fig:comparison_groundtruth}, we show the distance of the estimated distribution from the ground truth  in terms of the KL (Kullback-Leibler) metric, that captures the distance of the 2 distributions accounting for the bulk of the distributions. We also calculated the Kolmogorov-Smirnov (KS) statistic, not shown here, which captures the maximum vertical distance of two distributions. We found that RWRW is more efficient than MHRW with respect to both statistics.
We note that the usage of distance metrics such as KL and KS cannot replace the role of the formal diagnostics which are able to determine convergence online and most importantly in the absence of the ground truth.

\begin{figure}
\centering
\psfrag{RWRW}[cc][cc][0.4]{\hspace{-1.5mm} RWRW}
\psfrag{RWRW-Fair}[cc][cc][0.4]{RWRW-Uniq}
\psfrag{MHRW}[cc][cc][0.4]{\hspace{-1.5mm} MHRW}
\psfrag{MHRW-Fair}[cc][cc][0.4]{MHRW-Uniq}
\includegraphics[scale=0.23]{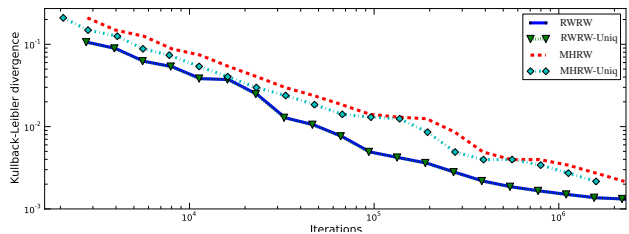}  %
\caption{The efficiency of RWRW and MHRW in estimating the degree distribution of \facebook, in terms of the Kullback-Leibler (KL) divergence. 
The ``Uniq'' plots count as iterations only the number of unique sampled nodes, which represents the real bandwidth cost of sampling. 
}
\label{fig:comparison_groundtruth}
\end{figure}

\subsubsection{\label{sec:fb_metrics}The choice of metric matters}
\label{subsec:fb_Dependence on the metric}

MCMC is typically used to estimate some user property/metric, {\em i.e.,} a function of the underlying random variable. The choice of metric can greatly affect the convergence time. We chose the metrics in the diagnostics, guided by the following principles:
\begin{itemize}
\item We chose the node degree because it is one of the metrics we want to estimate; therefore we need to ensure that the MCMC has converged at least with respect to it. The distribution of the node degree is also typically heavy tailed, and thus is slow to converge.
\item We also used several additional metrics (e.g. network ID, user ID and membership to specific networks), which are uncorrelated to the node degree and to each other, and thus provide additional assurance for convergence. 
\item We essentially chose to use all the nodal attributes that were easily and cheaply accessible at each node. These metrics were also relevant to the estimation at later sections. 
\end{itemize}

\begin{figure*}
\centering
\psfrag{Prob(deg)}[c][t][0.9]{$\Prob(k_v=k)$}
\psfrag{node degree}[c][b][0.9]{node degree $k$}
\psfrag{Prob(region)}[c][t][0.9]{$\Prob(N_v=N)$}
\psfrag{regional network}[c][b][0.9]{regional network $N$}
\includegraphics[width=0.9\textwidth]{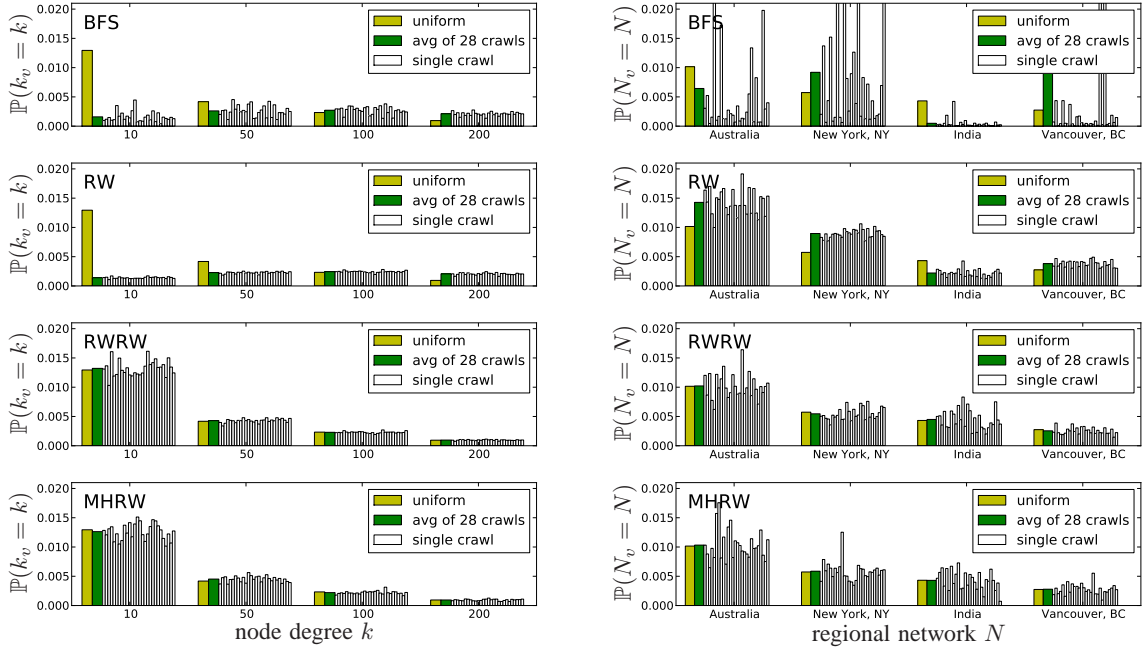}
\caption{Histograms of visits at node of a specific degree (left) and in a specific regional network (right). We consider four sampling techniques: BFS, RW, RWRW and MHRW.
We present how often a specific type of node is visited by the 28 crawlers (`crawls'), and by the uniform UNI sampler (`uniform'). We also plot the visit frequency averaged over all the 28 crawlers (`average crawl'). Finally, `size' represents the real size of each regional network normalized by the the total \facebook size.
\quad We used all the 81K nodes visited by each crawl, except the first 6K burn-in nodes. The metrics of interest cover roughly the same number of nodes (about 0.1\% to 1\%), which allows for a fair comparison.}
\label{fig:histograms_MHRW_BFS_RW}
\end{figure*}

Let us focus on two of these metrics of interest, namely {\em node degree}  and {\em sizes of geographical network} and study their convergence in more detail. The results for both metrics and all three methods are shown in Fig.\ref{fig:histograms_MHRW_BFS_RW}. We expected node degrees to not depend strongly on geography, while the relative size of geographical networks to strongly depend on geography. This implies that (i) the degree distribution will converge fast to a good uniform sample even if the walk has poor mixing and stays in the same region for a long time; (ii) a walk that mixes poorly will take long time to barely reach the networks of interest, not to mention producing a reliable network size estimate.
In the latter case,  MHRW will need a large number of iterations before collecting a representative sample.

The results presented in Fig.~\ref{fig:histograms_MHRW_BFS_RW}~(bottom) confirm our expectation. MHRW performs much better when estimating the probability of a node having a given degree, than the probability of a node belonging to a specific regional network. For example, one MHRW crawl overestimates the size of ``New York, NY'' by roughly 100\%. The probability that a perfect uniform sampling makes such an error (or larger) is $\sum_{i=i_0}^\infty {i \choose n}p^i(1-p)^i\simeq 4.3\cdot 10^{-13}$, where we took $i_0=1k$, $n=81K$ and $p=0.006$. Even given such single-walk deviations, the multiple-walk average for the MHRW crawl provides an excellent estimate of the true population size.

\subsection{\label{sec:fb_uniformity}Unbiased Estimation}

This section presents the main results of this chapter. First, the MHRW and RWRW methods perform very well: they estimate two distributions of interest (namely node degree, regional network size) essentially identically to the UNI sampler. Second, the baseline algorithms (BFS and RW) deviate substantively from the truth and lead to misleading estimates.

\subsubsection{Node degree distribution}

\begin{figure}%
\centering
\psfrag{Prob(deg=x)}[c][t][0.9]{$\Prob(k_v=k)$}
\psfrag{Prob(deg>=x)}[c][t][0.9]{$\Prob(k_v\geq k)$}
\psfrag{x}[c][b][0.75]{node degree $k$}
\includegraphics[width=0.5\textwidth]{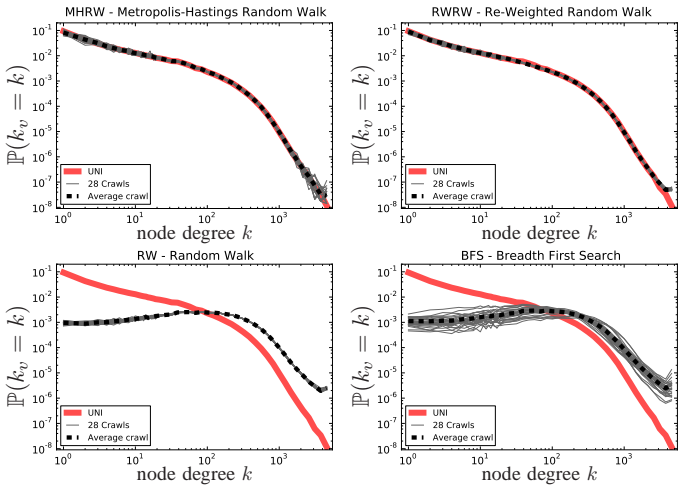}
\caption{Degree distribution (pdf) estimated by the sampling techniques  and the ground truth (uniform sampler). MHRW and RWRW agree almost perfectly with the UNI sample; while BFS and RW deviate significantly. We use log-log scale and logarithmic binning of data in all plots.}
\label{fig:degree_PDFs_loglog}
\end{figure}

In Fig.~\ref{fig:degree_PDFs_loglog} we present the degree distributions estimated by MHRW, RWRW,  RW, and BFS. The average MHRW crawl's pdf, shown in Fig.~\ref{fig:degree_PDFs_loglog}(a) is virtually identical to UNI. Moreover, the degree distribution found by each of the 28 chains separately are almost identical. In contrast, RW and BFS shown in Fig.~\ref{fig:degree_PDFs_loglog}(c) and (d) introduce a strong bias towards the high degree nodes. For example, the low-degree nodes are under-represented by two orders of magnitude. As a result, the estimated average node degree is $\overline{k}_v\simeq95$ for MHRW and UNI, and $\overline{k}_v\simeq330$ for BFS and RW. Interestingly, this bias is almost the same in the case of BFS and RW, but BFS is characterized by a much higher variance.
Notice that that BFS and RW estimate wrong not only the parameters but also the shape of the degree distribution, thus leading to wrong information.
Re-weighting the simple RW corrects for the bias and results to RWRW, which performs almost identical to UNI, as shown in \ref{fig:degree_PDFs_loglog}(b). As a side observation we can also see that the true degree distribution clearly {\em does not} follow a power-law.

\subsubsection{Regional networks}
Let us now consider a geography-dependent sensitive metric, {\em i.e.,} the relative size of regional networks. The results are presented in Fig.~\ref{fig:histograms_MHRW_BFS_RW}~(right). BFS performs very poorly, which is expected due to its local coverage. RW also produces biased results, possibly because of a slight positive correlation that we observed between network size and average node degree. In contrast, MHRW and RWRW perform very well albeit with higher variance, as already discussed in Section~\ref{subsec:fb_Dependence on the metric}.

\begin{figure}
\centering
\psfrag{degree}[c][b][0.7]{degree $k_v$}
\psfrag{nodeID2}[c][c][0.7]{userID}
\psfrag{nodeID}[c][b][0.9]{userID}
\psfrag{cdf}[c][t][0.9]{cdf}
\psfrag{Prob(deg>=x)}[c][t][0.9]{$\Prob(k_v\geq k)$}
\includegraphics[width=0.40\textwidth]{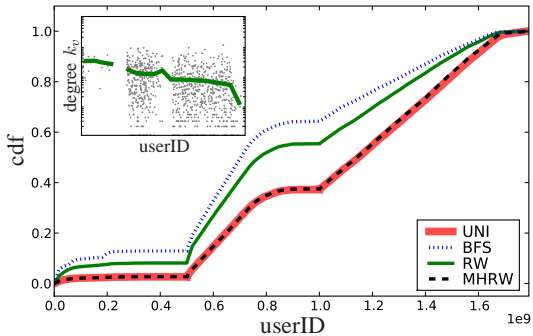}
\caption{User ID space usage discovered by BFS, RW, MHRW and UNI. Each user is assigned a 32 bit long userID. Although this results in numbers up to $4.3e9$, the values above $1.8e9$ almost never occur. \quad Inset: The average node degree (in log scale) as a function of userID.}
\label{fig:userIDspace}
\end{figure}

\subsubsection{The userID space}
\label{subsection:userID}

Finally, we look at the distribution of a property that is completely uncorrelated from the topology of \facebook, namely the user ID.  When a new user joins \facebook, it is automatically assigned a 32-bit number, called userID. It happens before the user specifies its profile, joins networks or adds friends, and therefore one could expect no correlations between userID and these features. In other words, the degree bias of BFS and RW should not affect the usage of userID space. Therefore, at first,  we were surprised to find big differences in the usage of userID space discovered by BFS, RW and MHRW. We present the results in Fig~\ref{fig:userIDspace}.

Note that the userID space is not covered uniformly, probably for historical reasons.
BFS and RW are clearly shifted towards lower userIDs. The origin of this shift is probably historical. The sharp steps at $2^{29}\!\!\simeq\!\!0.5\textrm{e}9$ and at $2^{30}\!\!\simeq\!\!1.0\textrm{e}9$ suggest that \facebook was first using only 29 bit of userIDs, then 30, and now 31. As a result, users that joined earlier have the smaller userIDs. At the same time, older users should have higher degrees on average, which implies that userIDs should be negatively correlated with node degrees. This is indeed the case, as we show in the inset of Fig~\ref{fig:userIDspace}.\footnote{Our observations are also confirmed by internal sources within \facebook \cite{facebook_uidspacehistory}. According to them, \facebook's user ID assignment reflects the history of the website and has transitioned through several phases. Initially, userIDs were auto-incremented starting at 4. As more networks, such as colleges or high schools, were supported, customized userID spaces were assigned per college \ie  Stanford IDs were assigned between 200000-299999. Finally, open registration to all users introduced scalability problems and made userID assignment less predictable.}
 This, together with the degree bias of BFS and RW, explains the shifts of userIDs distributions observed in the main plot in Fig~\ref{fig:userIDspace}. In contrast to BFS and RW,  MHRW performed extremely well with respect to the userID metric.

\subsection{\label{sec:fb_recommendation}Findings and Practical Recommendations}

\subsubsection{Choosing between methods} First and most important, the above comparison demonstrates that both MHRW and RWRW succeed in estimating several \facebook properties of interest virtually identically to UNI. In contrast,  commonly used baseline methods (BFS and simple RW) fail, {\em i.e.,} deviate significantly from the truth and lead to substantively erroneous estimates. Moreover, the bias of BFS and RW shows up not only when estimating directly node degrees (which was expected), but also when we consider other metrics seemingly uncorrelated metrics (such as the size of regional network), which end up being correlated to node degree. This makes the case for moving from ``1st generation'' traversal methods such as BFS, which have been predominantly used in the measurements community so far \cite{Ahn-WWW-07, Mislove2007, Wilson09}, to more principled, ``2nd generation'',  sampling techniques whose bias can be analyzed and/or corrected for. The random walks considered in this paper, RW, RWRW and MHRW, are well-known in the field of Monte Carlo Markov Chains (MCMC). We apply and adapt these methods to \facebook, for the first time, and we demonstrate that, when appropriately used, they perform remarkably well on real-world OSNs.

\subsubsection{Adding convergence diagnostics and parallel crawls} A key ingredient of our implementation - to the best of our knowledge, not previously employed in network sampling - was the use of formal {\em online} convergence diagnostic tests. We tested these on several metrics of interest within and across chains, showing that convergence was obtained within a reasonable number of iterations. %
We believe that such tests can and should be used in field implementations of walk-based sampling methods to ensure that samples are adequate for subsequent analysis. Another key ingredient of our implementation was the use of independent crawlers (started from several random independent starting points, unlike \cite{Rasti09-RDS, Rasti2008} who use a single starting point), which both improved convergence and decreased the duration of the crawls.

\subsubsection{ MHRW vs. RWRW} Both MHRW and RWRW achieved a uniform sample. When comparing the two, RWRW is
slightly more efficient, \ie needed less samples for the same accuracy. This is consistent
with the findings in \cite{Rasti09-RDS, Rasti2008}. This is partly due to the fact that MHRW requires a large
number of rejections during the initial sampling process; and partly due to slower mixing, in practice, as it avoids high degree nodes.
In this section, we present an empirical comparison based on the \facebook experiments. In Appendix C, we provide a more in-depth comparison via analysis and simulation.

However,
when choosing between the two methods there are additional trade-offs to
consider.  First, MHRW yields an asymptotically uniform sample, which
requires no additional processing for subsequent analysis.  By contrast,
RWRW samples are heavily biased towards high-degree nodes, and require
use of appropriate re-weighting procedures to generate correct results.
  For the creation of large data sets intended for general distribution
(as in the case of our \facebook sample), this ``ready-to-use'' aspect of
MHRW has obvious merit\footnote{For example, our released data sets \cite{our-dataset}
are intended to be used by people that are not necessarily experts in the re-weighting
methods, for whom the potential for erroneous misuse is high}.  A second
advantage of MHRW is the ease of online testing for convergence to the
desired target (uniform) distribution.  In contrast, in RWRW, we test for
convergence on a different distribution and then re-weight, which can
introduce distortion\footnote{
It is in principle possible to diagnose
convergence on re-weighted statistics with RWRW.  However, this requires appropriate use of
re-weighting during the convergence evaluation process, which can
increase the complexity of implementation.
}.
Finally, simple
re-weighting is difficult or impossible to apply in the context of many
purely data-analytic procedures such as multidimensional scaling or
hierarchical clustering. %
Simulated Importance Resampling \cite{Rubin:SIR} provides a useful
alternative for RWRW samples, but suffers from well-known problems of
asymptotic bias (see \cite{Skare03} for a discussion).  This is of less concern for applications such as moment
estimation, for which re-weighting is both simple and effective.

Ultimately, the choice of RWRW versus MHRW is thus a trade-off between
efficiency during the initial sampling process (which favors RWRW in all practical cases)
and simplicity/versatility of use for the resulting data set (which
favors MHRW).  For our present purposes, these trade-offs favor
MHRW, and we employ it here for producing a uniform ready-to-use sample of users.

\section{Facebook Characterization}
\label{sec:fb_characterization}

In this section, we use the uniform sample of 1M nodes, collected through MHRW, and the sub-sample of 37K extended egonets to study some features of \facebook. In contrast to previous work, which focused on particular regions \cite{harvard-dataset, caltech-dataset} or used larger but potentially biased samples \cite{Mislove2007, Wilson09}, our user sample is representative of the entire \facebook.

\subsection{Topological Characteristics}
\label{sec:fb_characterization_topological}

We first focus on purely topological aspects of the graph of \facebook.

\subsubsection{Degree distribution}
Node degree is just one example of a node-level (user) property, which also happens to be an important topological characteristic
In Fig.~\ref{fig:degree_PDFs_loglog}(a,b), we present the node degree distribution of \facebook.
 Differently from previous studies of crawled datasets in online social networks in \cite{Wilson09, Mislove2007, Ahn-WWW-07}, we observe that node degree \emph{is not} a power law. Instead, we can identify two regimes, roughly $1\!\leq\!k\!<\!300$ and $300\!\leq\!k\!\leq\!5000$, each of which can be approximated by a power-law with exponents $\alpha_{k<300}=1.32$ and $\alpha_{k\geq300}=3.38$, respectively.
We note, however, that the regime $300 \leq k \leq 5000$ covers only slightly more than one decade. This behavior is suggestive of
multistage ``vetting" models of network formation.

\begin{figure}[t!]
\centering
\psfrag{neighbor node degree}[c][t][0.9]{neighbor node degree $k'$}
\psfrag{user node degree}[c][b][0.9]{node degree $k$}
\includegraphics[width=0.4\textwidth]{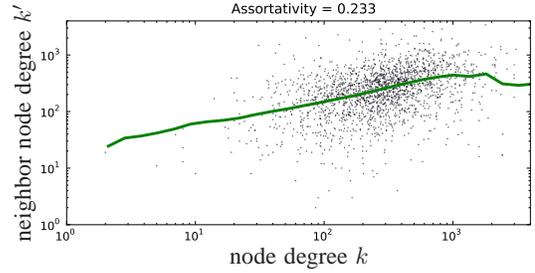}
\caption{Assortativity - correlation between degrees of neighboring nodes. The dots represent the degree-degree pairs (randomly subsampled for visualization only).  The line uses log-binning and takes the average degree of all nodes that fall in a corresponding bin.
}
\label{fig:Assortativity}
\end{figure}

\begin{figure}[t!]
\centering
\psfrag{Clustering Coefficient}[c][t][0.9]{clustering coefficient $C(k)$}
\psfrag{node degree}[c][b][0.9]{node degree $k$}
\includegraphics[width=0.4\textwidth]{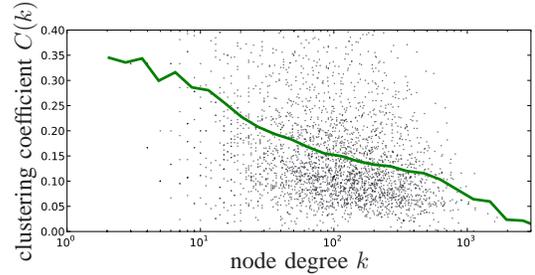}
\caption{Clustering coefficient of \facebook users as function of their degree.}
\label{fig:ClusteringCoefficient}
\end{figure}

\subsubsection{Assortativity} Depending on the type of complex network, nodes may tend to connect to similar or different nodes. For example, in many social networks high degree nodes tend to connect to other high degree nodes~\cite{Newman02}. Such networks are called \emph{assortative}.
In Fig.\ref{fig:Assortativity}, we plot the node degree vs. the degrees of its neighbors. We observe a positive correlation, which implies assortative mixing and is in agreement with previous studies of similar social networks. We can also summarize this plot by calculating the Pearson correlation coefficient, or \emph{assortativity coefficient} which is $r=0.233$. This value is higher than $r'=0.17$ reported in \cite{Wilson09}. A possible explanation is that the Region-Constrained BFS used in \cite{Wilson09} stops at regional network boundaries and thus misses many connections to, typically high-degree, nodes outside the network.

\subsubsection{Clustering coefficient}
In social networks, it is likely that two friends of a user are also friends of each other. The intensity of this phenomenon can be captured by the \emph{clustering coefficient} $C_v$ of a node $v$, defined as the relative number of connections between the nearest neighbors of~$v$.
The clustering coefficient of a network is the  average $C$ over all nodes. %
We find the average clustering coefficient of \facebook  to be $C=0.16$, similar to that reported in \cite{Wilson09}.
Since the clustering coefficient tends to depend strongly on the node's degree~$k_v$, we looked at $C_v$ as a function of $k_v$.
Fig.~\ref{fig:ClusteringCoefficient} shows
a larger range in the degree-dependent clustering coefficient ([0.05, 0.35]) than what was found in \cite{Wilson09} ([0.05, 0.18]).

\subsection{Privacy awareness}

 \begin{figure}%
\centering
 \psfrag{Prob}[c][t][0.8]{$\Prob(Q_v)$}
 \psfrag{Privacy settings}[c][b][0.9]{privacy settings $Q_v$}
 \includegraphics[width=0.48\textwidth]{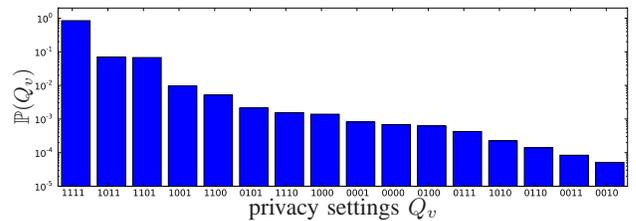}
 \caption{The distribution of the privacy settings among $\sim$ 172M \facebook users.  Value $Q_v\!=\!1111$ corresponds to default settings (privacy not restricted) and covers 84\% of all users.}
 \label{fig:PrivacySettings_hist}
 \end{figure}

We also collected the privacy settings $Q_v$ for each node $v$. $Q_v$ consists of four bits, each corresponding to one privacy attribute. By default, \facebook sets these attributes to `allow', i.e., $Q_v\!=\!1111$ for a new node $v$. We call users who change these default settings as `privacy aware' users, and we denote by $PA$ the level of privacy awareness of a user $v$, {\em i.e.,} privacy aware users have $PA=\Prob(Q_v\!\neq\!1111)$.
\begin{table}
\centering
 {\small
 \begin{tabular}{|l|l|l|l|}
 \hline
 {\footnotesize $PA$}   & {\footnotesize Network $n$}  & {\footnotesize $PA$}   & {\footnotesize Network $n$}\\
 \hline
 {\footnotesize 0.08}	& {\footnotesize Iceland}	& {\footnotesize  \ldots} & {\footnotesize \ldots }\\
 {\footnotesize 0.11}	& {\footnotesize Denmark}	& {\footnotesize 0.22}	& {\footnotesize Bangladesh} \\
 {\footnotesize 0.11}	& {\footnotesize Provo, UT}	& {\footnotesize 0.23}	& {\footnotesize Hamilton, ON}\\
 {\footnotesize 0.11}	& {\footnotesize Ogden, UT} 	& {\footnotesize 0.23}	& {\footnotesize Calgary, AB}\\
 {\footnotesize 0.11}	& {\footnotesize Slovakia}	& {\footnotesize 0.23}	& {\footnotesize Iran}\\
 {\footnotesize 0.11}	& {\footnotesize Plymouth}	& {\footnotesize 0.23}	& {\footnotesize India}\\	
 {\footnotesize 0.11}	& {\footnotesize Eastern Idaho, ID} & {\footnotesize 0.23}	 & {\footnotesize Egypt}\\
 {\footnotesize 0.11}	& {\footnotesize Indonesia}	& {\footnotesize 0.24}		& {\footnotesize United Arab Emirates}\\
 {\footnotesize 0.11}	& {\footnotesize Western Colorado, CO} & {\footnotesize 0.24}	& {\footnotesize Palestine}\\
 {\footnotesize 0.11}	& {\footnotesize Quebec City, QC}	& {\footnotesize 0.25}	& {\footnotesize Vancouver, BC}\\
 {\footnotesize 0.11}	& {\footnotesize Salt Lake City, UT}	& {\footnotesize 0.26}	& {\footnotesize Lebanon}\\
 {\footnotesize 0.12}	& {\footnotesize Northern Colorado, CO} 	& {\footnotesize 0.27}	& {\footnotesize Turkey}\\
 {\footnotesize 0.12}	& {\footnotesize Lancaster, PA}	& {\footnotesize 0.27}	& {\footnotesize Toronto, ON}\\
 {\footnotesize 0.12}	& {\footnotesize Boise, ID}	& {\footnotesize 0.28}	& {\footnotesize Kuwait}\\
 {\footnotesize 0.12}	& {\footnotesize Portsmouth}	& {\footnotesize 0.29}	& {\footnotesize Jordan}\\
 {\footnotesize \ldots} & {\footnotesize \ldots} & {\footnotesize 0.30}	& {\footnotesize Saudi Arabia} \\
   \hline
 \end{tabular}
 }
 \caption{{\footnotesize Regional networks with respect to their privacy awareness $PA =\Prob(Q_v\!\!\neq\!\!1111\ | v\!\in\!n)$ among $\sim$ 172M \facebook users. Only regions with at least 50K users are considered. Note the different types of countries in the two extreme ends of the spectrum. {\em E.g.} many \facebook users in the Middle East seem to be highly concerned about privacy. Scandinavian users are the least privacy concerned. Canada regions show up at both ends, clearly splitting into English and French speaking parts.}}
 \label{tab:Privacy_vs_network}
 \end{table}

 We studied the privacy awareness in \facebook and we report some of our findings.
 First, we present the distribution of privacy settings among \facebook users in Fig.~\ref{fig:PrivacySettings_hist}, which shows that about 84\% of users leave the settings unchanged, {\em i.e.,} $\Prob(Q_v\!\!=\!\!1111)\simeq0.84$. The remaining 16\% of users modified he default settings, yielding  $PA=0.16$ across the entire \facebook. The two most popular modifications are $Q_v\!=\!1011$ (``hide my photo'') and $Q_v\!=\!1101$ (``hide my friends''), each applied by about 7\% of users.
 Second, the privacy awareness $PA$ of \facebook users depends on many factors, such as the geographical location, node degree or the privacy awareness of friends. In Table~\ref{tab:Privacy_vs_network} we classify the regional networks with respect to $PA$ of their members.
 \begin{figure}%
\centering
 \psfrag{Prob(not 1111)}[c][t][0.9]{$PA$ - privacy awareness}
 \psfrag{node degree}[c][b][0.8]{node degree $k$}
 \includegraphics[width=0.48\textwidth]{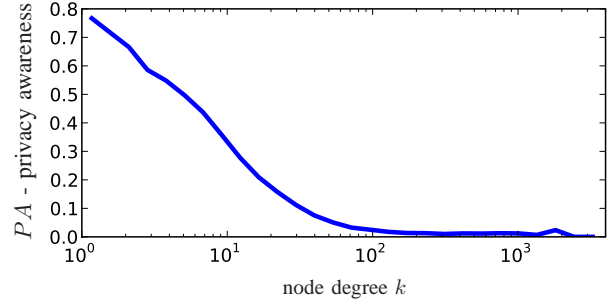}
\caption{Privacy awareness as a function of node degree in the egonets dataset. We consider only the nodes with privacy settings set to '**1*', because only these nodes allow us to see their friends and thus degree. So here $PA = \Prob(Q_v\!\!\neq\!\!1111\ |\ k_v = k,\ Q_v\!\!=\!**1*)$.}
 \label{fig:privacy_vs_deg}
 \end{figure}
 In Fig.~\ref{fig:privacy_vs_deg}, we show the effect of node degree on  the privacy settings of a user.
 We found that low degree nodes tend to be very concerned about privacy, whereas high degree nodes hardly ever bother to modify the default settings. This clear trend makes sense in practice: to protect her privacy, a privacy concerned user would carefully select her \facebook friends, {\em e.g.,} by avoiding linking to strangers. At the other extreme, there are users who prefer to have as many `friends' as possible, which is much easier with unrestricted privacy attributes.
 \begin{figure}%
\centering
 \psfrag{Prob(not 1111)}[c][t][0.8]{$PA$ - privacy awareness}
 \psfrag{average neighbor privacy concern}[c][b][0.9]{$\overline{PA}$ - average privacy awareness of node's neighbors}
 \includegraphics[width=0.48\textwidth]{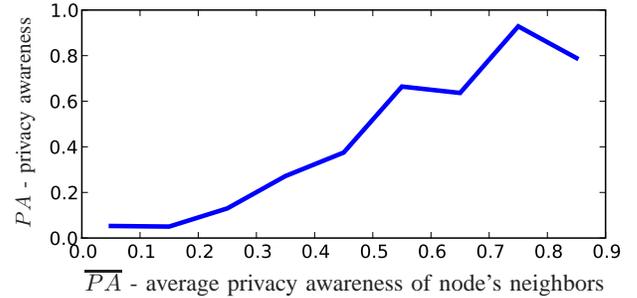}
 \caption{Privacy awareness  as a function of privacy awareness of node's neighbors in the egonets dataset. We consider only the nodes with privacy settings set to '**1*', because only these nodes allow us to see their friends, so $PA = \Prob(Q_v\!\!\neq\!\!1111\ |\ \overline{PA},\ Q_v\!\!=\!**1*)$.}
 \label{fig:privacy_vs_privacy}
 \end{figure}
 Finally, we found that the privacy awareness of a user is positively correlated with the privacy awareness of her friends.
  We observe a clear positive correlation in Fig.~\ref{fig:privacy_vs_privacy}.

\section{Conclusion}
\label{sec:conclude}

In this paper, we developed a framework for unbiased sampling of users in an OSN by crawling the social graph, and we provided recommendations for its implementation in practice. We made the following contributions.
We compared several candidate techniques in terms of bias (BFS and RW were significantly biased, while MHRW and RWRW provided unbiased samples) and efficiency (we found RWRW to be the most efficient in practice, while MHRW has the advantage of providing a ready-to-use sample). We also introduced the use of formal online convergence diagnostics. In addition, we performed an offline comparison of all crawling methods against the ground truth (obtained through uniform sampling of userIDs via rejection sampling). 
We also provided guidelines for implementing high performance crawlers for sampling OSNs.
Finally, we applied these methods to \facebook and obtained the first unbiased sample of \facebook users, which we used it to characterize several key user and structural  properties of \facebook. We anonymized our datasets and made them publicly available at \cite{our-dataset}.  They have been downloaded approximately 500 times at the time of this publication.

\section*{Appendix A: Uniform Sample of User IDs with Rejection Sampling}
\label{sec:appendix_UNIsampling}

In Section \ref{sec:fb_uni} we obtained  an exact uniform sample  of \facebook users by directly querying the userID space. We then used it as ground truth for comparing against the samples collected by the crawling techniques.
More specifically, we used the following procedure:

{\bf UNI Sampling Method:} First, pick a userID $X$ uniformly at random in the known range of IDs [0,{\tt MAXuserID}].
Query the OSN for user data for the chosen userID $X$. If the OSN does not return an error message (\ie $X$ is valid/allocated), include  $X$ in the sample; otherwise discard it.
Repeat the process.
\hfill $\blacksquare$

Note that the userID space is {\em not} required to be sequentially or evenly allocated. In fact, as we explicitly show in Section \ref{subsection:userID} and Fig.\ref{fig:userIDspace}, the \facebook userID space is {\em} not allocated in a sequential way. Nevertheless, by using rejection sampling, we are guaranteed to obtain a uniform sample of the {\em allocated} userIDs (not of all userIDs), as the following proposition indicates.

{\bf Proposition:}
UNI yields a uniform sample of the {\em allocated} user IDs in an Online Social Network.

{\em Proof.} Let us consider that there are $N$ allocated and $M$ non-allocated userIDs in the entire userID space [0,{\tt MAXuserID}]. These $N$ and $M$  userIDs do not need to be consecutive. UNI picks any element, w.p. $\frac{1}{N+M}$ and accepts it if it is valid (w.p. $Pr\{accepted\}=\frac{N}{N+M}$); otherwise it rejects it (w.p. $\frac{M}{N+M}$). It is easy to see the distribution of the accepted  userIDs is uniform:
$$Pr\{X|accepted\}=\frac{1}{N}.$$
Indeed, $Pr\{X|accepted\}=\frac{Pr\{X \textit{~and~accepted}\}}{Pr\{\textit{accepted}\}}$. If the userID is valid then $Pr\{X|\textit{accepted}\}=\frac{\frac{1}{N+M}\cdot 1}{\frac{N}{N+M}}=\frac{1}{N}$; otherwise it is 0.
\hfill $\blacksquare$

The above is just a special case of textbook {\em rejection sampling} (\eg see \cite{leon-garcia}, Chapter 3), when the desired sampling distribution is uniform. It is only repeated here for completeness. A common misunderstanding is to interpret UNI as a uniform sample of the entire userID space, while it is a uniform sample only of the allocated/valid userIDs, independently from where in they userID space these IDs may be.

 {\bf Limitations.}  There are some requirements for being able to implement UNI, which are met in our measurements of \facebook.
 \begin{itemize}
 \item First, we need to know or estimate the range of userIDs assigned to users, or equivalently the maximum known userID, {\tt MAXuserID}. Since we use rejection sampling, overestimating {\tt MAXuserID} simply results to more rejections. Knowledge of the actual distribution of userIDs in the userID space is {\em not} needed, as explained above.
 \item Second, we need to be able to query the OSN to return data for the selected userID, if it is valid, or return an error message if it does not exist. This was supported by \facebook at the time of our experiment.
 \item Furthermore, UNI is efficient only if the probability of acceptance is high. This was the case at the time of our experiments, when the userID was 32bits, but is no longer the case now that \facebook switched to  64bit IDs.
\end{itemize}
We emphasize that we use UNI only as a baseline for comparing the crawling methods against ground truth.

\section*{Appendix B: Temporal Dynamics}

The \facebook friendship graph can be considered practically static in the timescales our crawls, consistently with our assumption A4. \facebook is growing in general (as reported by websites such as \cite{facebook-data,inside-bacebook}) but in much longer timescales than the duration of our crawls (which were in the order of a few days, thanks to the efficient crawlers we developed). To confirm that this is indeed the case in \facebook, we took the following steps.

\begin{figure}
\centering
\psfrag{Prob(deg=x)}[c][t][0.9]{$\Prob(k_v=k)$}
\psfrag{yprob}[c][t][0.9]{$\Prob(Q_v=Q)$}
\psfrag{x}[c][b][0.8]{node degree $k$}
\psfrag{priv}[c][b][0.8]{Privacy bits $Q$}
\subfigure[Node Degree]{
\includegraphics[width=.35\textwidth]{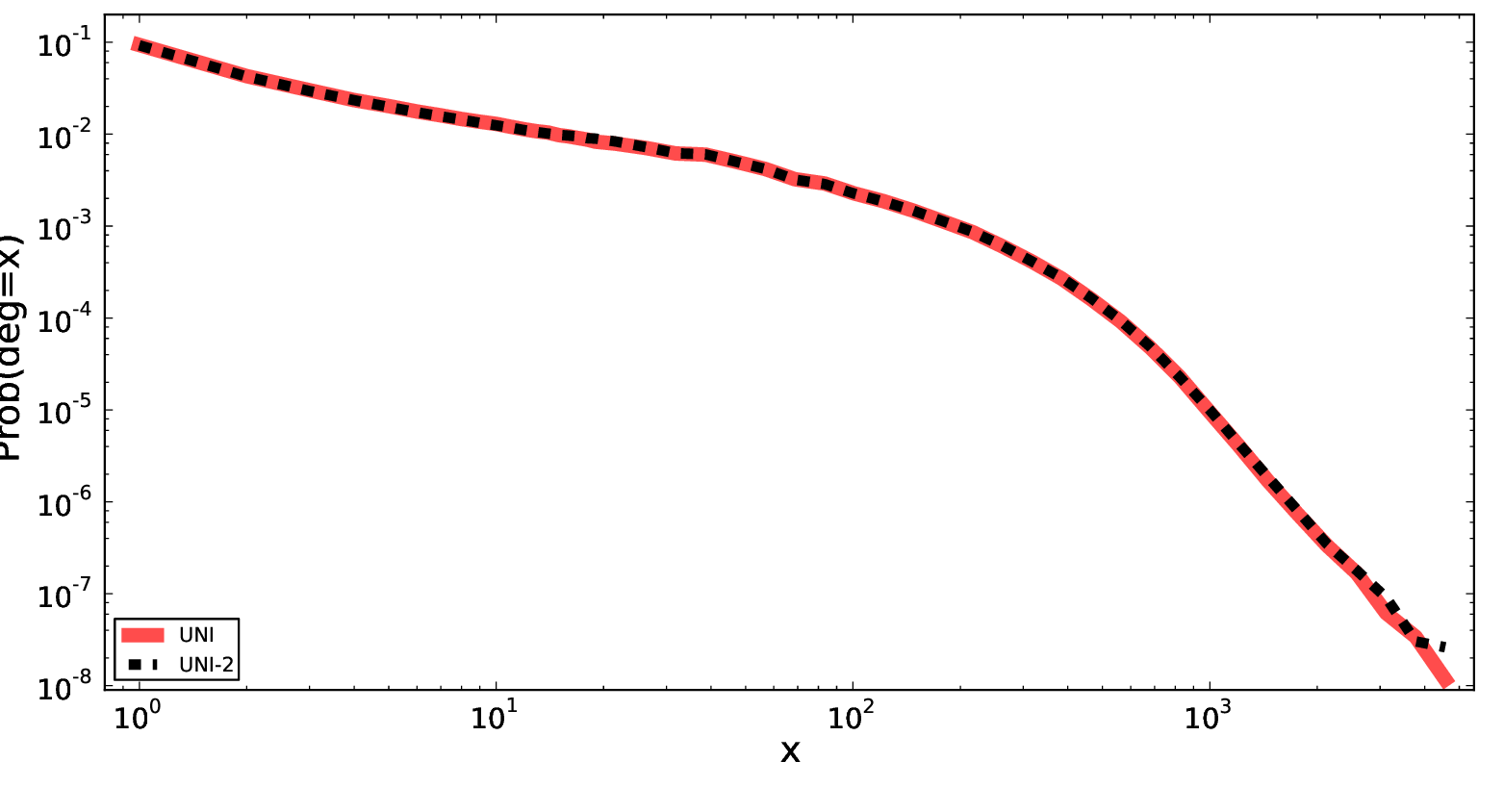}}
\subfigure[Privacy Settings $Q_v$]{
\includegraphics[width=.35\textwidth]{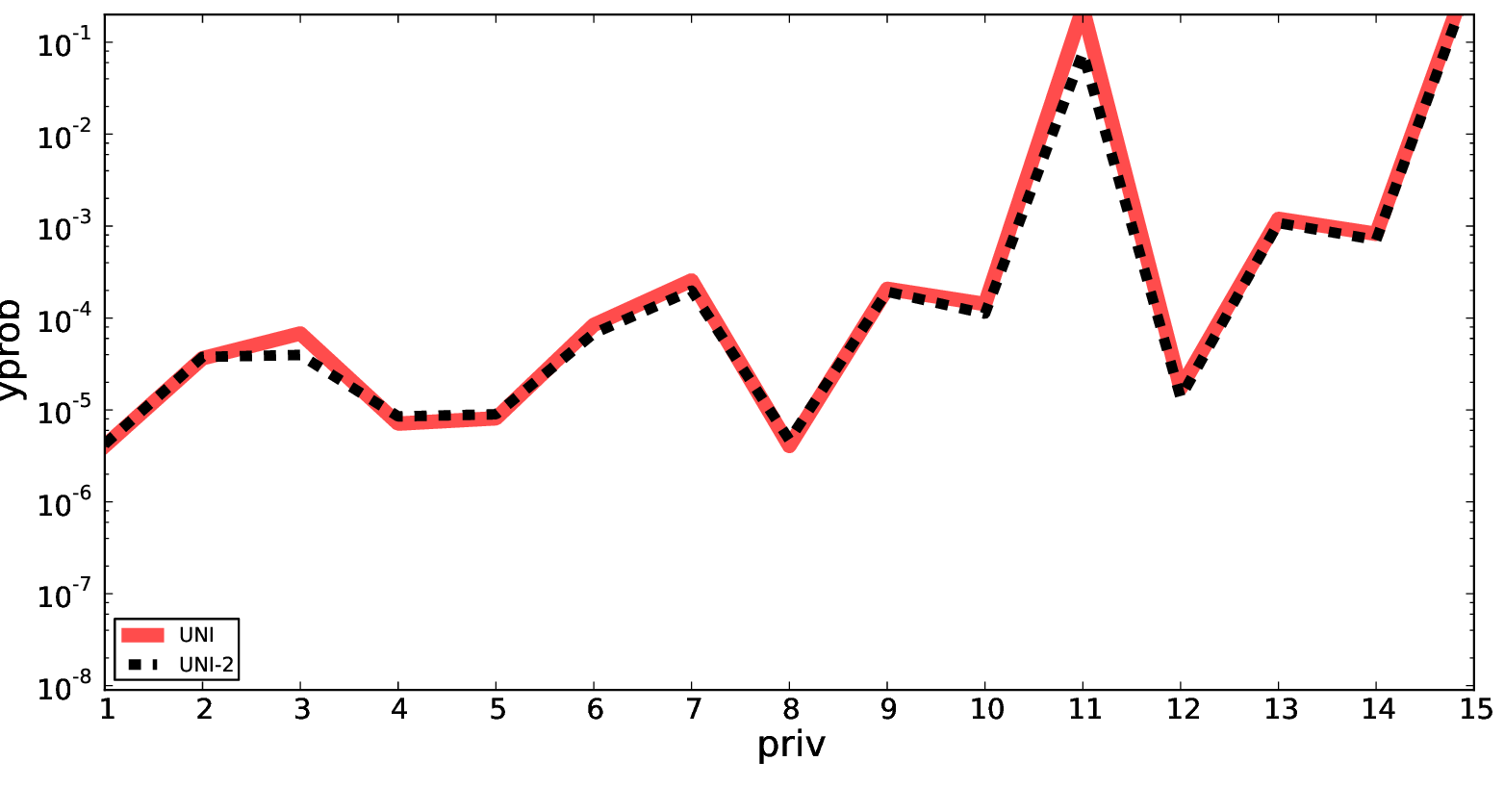}}
\caption{PDF of two user properties for UNI samples obtained 45 days apart}
\label{fig:unicomp}
\end{figure}

First, we compared our metrics of interest between the UNI sample of Table \ref{tab:datasets} and a similarly sized UNI sample obtained 45 days later. Fig~\ref{fig:unicomp} shows the distribution of node degree and privacy settings. One can easily observe that the two distributions we obtained were virtually identical. The same was true in all other comparisons we did on the same metrics of interest in the each sample across different days.

Second, during the period that we performed our crawls (see table \ref{tab:datasets}), \facebook was growing at a rate of $450K/day$ as reported by websites such as \cite{facebook-data,inside-bacebook}. With a population of $\sim$200M users during that period, this translates to a growth of 0.22\% of users/day.
Each of our crawls lasted around 4-7 days (during which, the total \facebook growth was 0.9\%-1.5\%); in fact, our convergence analysis shows that the process converged even faster, {\em i.e.,} in only one day. Therefore, the relative node growth of \facebook users was negligible during our crawls.

\begin{figure}
\centering
\psfrag{Dkv/kv}[c][t][0.55]{$\Delta k_{v}/k_v$}
\psfrag{Dkv}[c][t][0.55]{$\Delta k_{v}$}
\psfrag{kv}[c][t][0.55]{$k_{v}$}
\subfigure[Relative degree and absolute degree change as a function of node degree]{
\includegraphics[width=.47\textwidth]{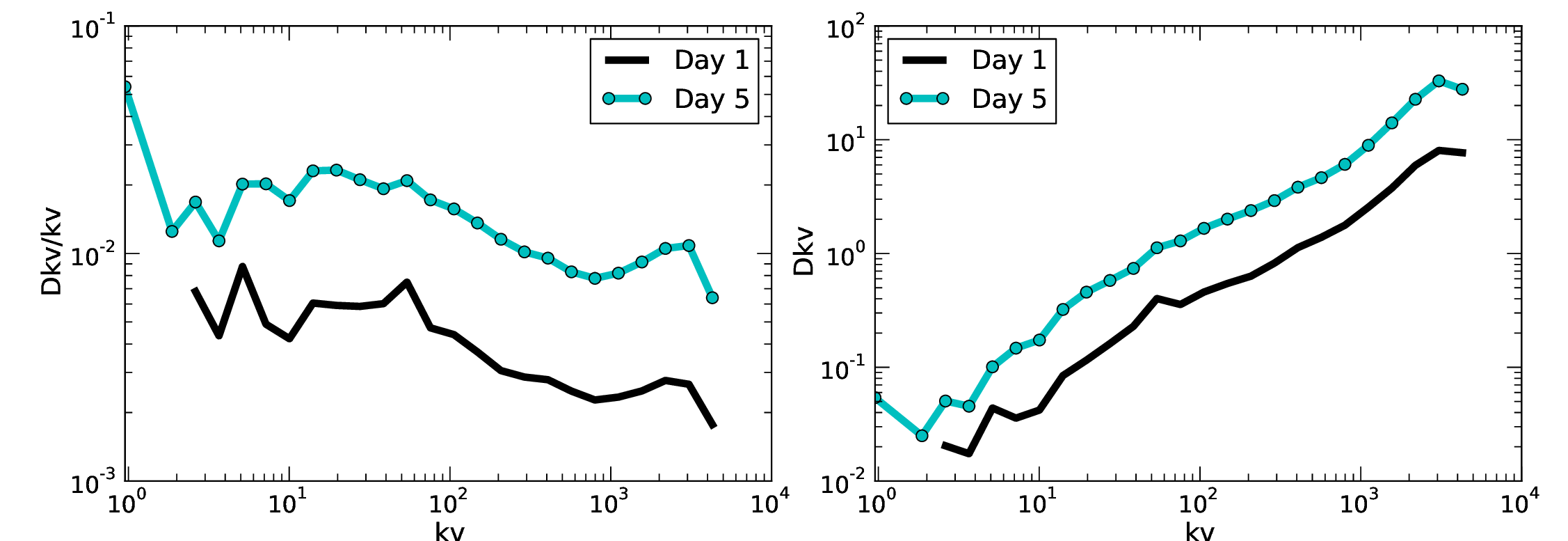}\label{fig:time_percentchange}}
\subfigure[Cumulative Distribution Function of relative degree and absolute degree change]{
\includegraphics[width=.47\textwidth]{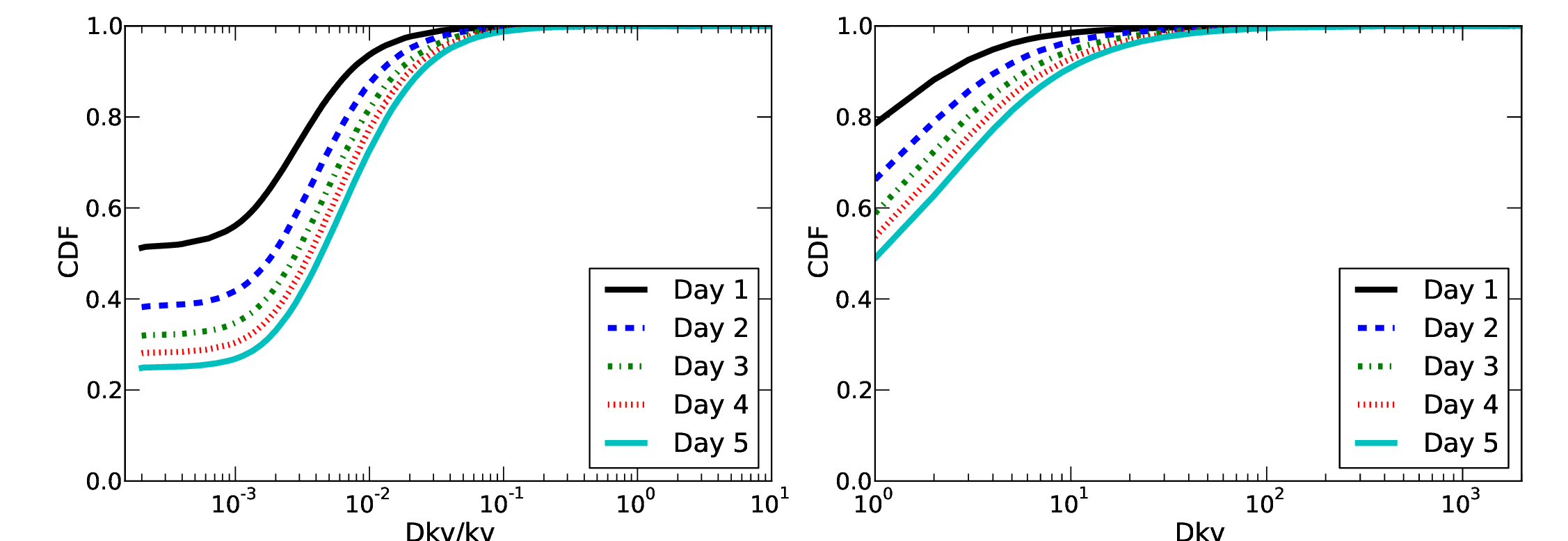}\label{fig:time_CDF}}
\caption{Daily growth of social graph for five days}
\label{fig:edgegrowth}
\end{figure}

Third, we evaluated the edge growth of the \facebook graph by conducting the following experiment. In November 2010 we ran 20 independent RW crawls and collected $20\times 10K = 200K$ user samples within 12 hours. This is our user sample $|S|^0$ at Day 0. For the next 5 days, we re-visit the exact same nodes in the same order and check their edges to see if there is any change. These will be our user samples  $|S|^1..|S|^5$, at Days 1,..,5 and their collection is approximately equally spaced in time. Let us define $\Delta k^{day}_{v} = |  k_v^{day} - k_v^0 | $ as the absolute degree change where $day \in [1..5]$ and $\Delta k^{day}_{v}/k_v^{day}$ as the relative degree change. Fig~\ref{fig:time_percentchange} shows the absolute degree and relative degree change as a function of node degree for Days 1,5. We observe that at Day 5 the relative degree change varies between 0.5\%-2\% depending on the node degree. In summary, the total relative change at Day 5 is estimated as  $$\frac{1}{|S|}\sum_{\substack{v \in |S|}}{\frac{\Delta k^{d5}_{v}}{k_v^{d5}}}$$
which is 1.13\%. Fig~\ref{fig:time_CDF} shows the CDF of relative and absolute degree change for Days 1..5. It can be seen that 50\% and 80\%  of the users at Day 5 and Day 1 respectively have an absolute degree change of one or zero. Additionally, 70\% and 95\%  of the users at Day 5 and Day 1 respectively have a relative degree change of less than 1\%.

Last, and perhaps most importantly, another way to evaluate the effect of edge growth in the estimation of the degree distribution (or any other property) is to compare $\frac{1}{|S|}\sum_{\substack{v \in |S|}}{({\Delta k^{d5}_{v}})^2}$ , which is  the node degree variance  due to temporal dynamics at day 5, to $\frac{1}{|S|}\sum_{\substack{v \in |S|}}{(\overline{k}-k_v^{d5})^2}$ , which is the variance of the node degree distribution at day 5. The former value is estimated at $556$ while the latter at $617,829$, which makes the node degree variance due to the temporal dynamics at least three orders of magnitude smaller than the distribution variance, \ie essentially noise.

These results are expected, since we deal with the social  graph, which is much more static than other types of graphs, \ie the interaction graph on OSNs or P2P graphs \cite{Stutzbach2006-unbiased-p2p, Willinger09-OSN_Research} that are known to have high churn. In the latter case, considering the dynamics becomes important. However, they appear not to be problematic for this particular study.

\section*{Appendix C: MHRW vs. RWRW}

 Although both MHRW and RWRW achieve asymptotically unbiased samples,  experimental results on \facebook in Fig.~\ref{fig:comparison_groundtruth} showed that RWRW outperforms MHRW in terms of sampling efficiency. %
This is consistent with what was reported by Rasti~\etal~\cite{Rasti09-RDS} in the context of unstructured peer-to-peer networks.
We investigate this further, by taking the following steps: 
(i)~we run simulations on a broad range of Internet topologies, 
(ii)~we attempt to gain more intuition via simplified analysis, and 
(iii)~we show a counterexample where MHRW is better than RWRW in a specific pathological case. 
Our conclusion is that RWRW is more efficient than MHRW in most topologies that are likely to arise in practice.

\subsection{Simulation results}

We compare RWRW with MHRW on a broad range of large, real-life, but fully known Internet topologies described in Table~\ref{tab:Real-life topologies}. As our main source of data we use the SNAP Graph Library \cite{WWW_SNAP_Graph_Library}. We show the results in Fig.~\ref{fig:MHRW_vs_RW_simulations}. The number on the bottom-left corner of every plot indicate how much longer MHRW should be than RWRW so as to achieve the same error.
One can see that, in all cases, MHRW requires 1.5-7 times more (unique) samples to achieve the same estimation quality as RWRW. Recall that in our \facebook measurement, this advantage of RWRW over MHRW is about 3.0 (see Fig.~\ref{fig:comparison_groundtruth}).

\subsection{Intuition}

Why does RWRW perform so much better than MHRW on real-life Internet topologies? There are three aspects that play a major role here, as discussed below.

\subsubsection{MHRW strongly re-samples low-degree nodes}
By design, MHRW prefers visiting low-degree nodes over high-degree ones, which compensates for the node degree bias observed under RW.
However, counter-intuitively, MHRW does not visit significantly more \emph{unique} low-degree nodes than RW does. Instead, it strongly re-samples the already-visited low-degree nodes. Indeed, under MHRW, the probability of leaving the node $u$ at any iteration is
\begin{equation}
	\label{eq:prob of leaving a node}
	p_u = \sum_{w \in\textrm{Neighbors of }u} \frac{1}{k_u} \cdot \min(1, \frac{k_u}{k_w})	
\end{equation}
Therefore, the number of rounds MHRW stays in node $u$ is a geometric random variable $Geom(p_u)$ with parameter $p_u$ and mean $1/p_u$.
In some graphs, $1/p_u$ may be in the order of thousands.

\subsubsection{Re-sampling geometric process introduces variance} %

RWRW compensates for the bias of RW by dividing each measured value of node~$u$ by its degree~$k_u$, which is a fixed number. In contrast, MHRW achieves it by re-sampling node~$u$, as shown above, which is a random process on its own that introduces additional variance. We demonstrate this using the example below.
Consider a star topology with node~$v^*$ in the middle and nodes~$V$ being its one-hop neighbors. 
Let $V'\subset V$ be the nodes of some type of interest. 
Our goal is to estimate the fraction $\theta=\frac{|V'|}{|V|}$.
To this end, we use RW and MHRW to collect a sample $S\subset V$ of nodes (for simplicity, we intentionally ignore node~$v^*$) of size $N=|S|$.
Clearly, we can estimate~$\theta$ by
$$\hat{\theta}\ =\ \frac{1}{N} \sum_{v\in S} 1_{v\in V'}.$$

RW alternates between $v^*$, and nodes chosen from~$V$ uniformly at random, with replacements. Therefore,
$$\hat{\theta}_{RW}\ =\ \frac{Binom(\theta, N)}{N} \quad \textrm{and}$$
$$\Var[\hat{\theta}_{RW}]\ =\ \frac{1}{N^2}\cdot \Var[Binom(\theta, N)]\ =\ \frac{\theta (1-\theta)}{N}.$$

In contrast, MHRW stays at any node from~$V$ for some number of rounds that follow a geometric distribution~$Geom(q)$. In our example, \eqn{\ref{eq:prob of leaving a node}} yields parameter~$q=1/|V|$. Therefore, MHRW spends on average $|V|$ rounds at every visited node from~$V$, which means that MHRW spends a large fraction of its $N$ samples (fraction $\frac{|V|-1}{|V|}$ on average) for re-sampling the same nodes,
which drastically limits the performance of the MHRW estimator.
In practice, however, re-sampling an already known node is free in terms of required bandwidth. Therefore, it makes sense to count all consecutive visits as 1 towards the sample size $N$, which yields
$$\hat{\theta}_{MHRW}\ =\ \frac{q}{N}  \sum_1^{Binom(\theta, N)} Geom(q). $$
Because $Binom(\theta_k, N)$ and $Geom(q)$ are independent, 
after some calculations, we get:
$$\Var[\hat{\theta}_{MHRW}]\ =\ \frac{\theta (1-\theta)}{N}+\frac{\theta (1-q)}{N}>\frac{\theta (1-\theta)}{N}= \Var[\hat{\theta}_{RW}].$$

\begin{table}[t!]
  \centering
\noindent {\footnotesize
\begin{tabular}{|r|r|r|l|}
\hline
    Dataset          & nodes   & edges &  Description \\		
\hline
    AS &      26K &      53K &  CAIDA AS Relationships Datasets, 2004\\
 Email &     225K &     341K &   Email network of a large Institution \cite{Leskovec2007}\\
   WWW &     326K &    1\,118K &  Web graph of Notre Dame, 1998 \cite{Albert-Nature-1999} \\
   P2P &      63K &     148K &   Gnutella p2p network, 08-2002 \cite{Leskovec2007}\\
   Slashdot &      77K &     546K &  Slashdot social network, 11-2008 \cite{Leskovec2009}\\
\hline
\end{tabular}}
  \caption{Real-life Internet topologies used in simulations. All graphs are connected and undirected (which required preprocessing in some cases).  }
  \label{tab:Real-life topologies}
\end{table}

\begin{figure}[t]
\centering
\psfrag{S}[c][c][0.8]{Sample length $|S|$}
\includegraphics[width=0.49\textwidth]{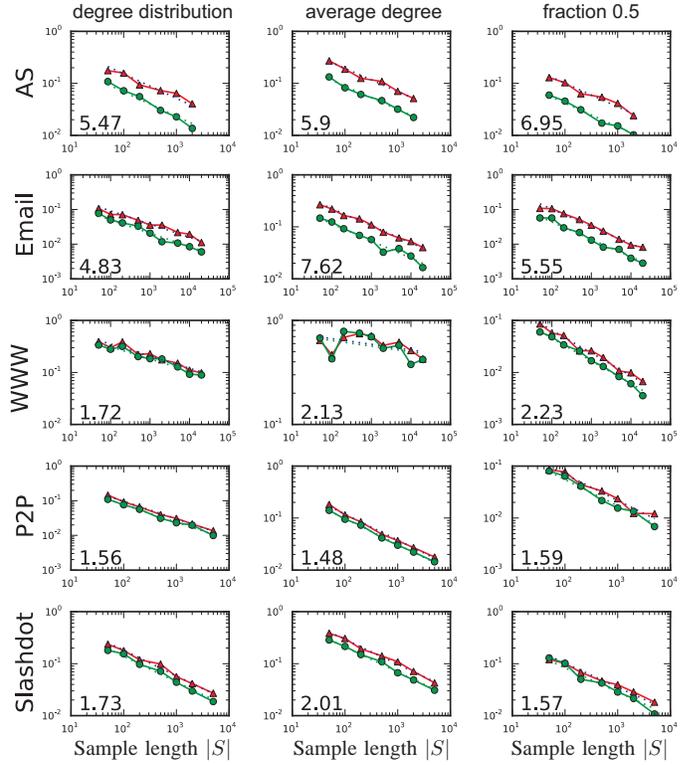}
\caption{Performance of MHRW and RWRW in simulations on Internet topologies described in Table~\ref{tab:Real-life topologies}.
We use MHRW and RWRW to collect samples. Based on 1000 such samples, we estimate three different graph parameters (below).
We draw the median estimation error of MHRW (red triangles) and RWRW (green circles) as a function of sample length~$|S|$ counted in \emph{unique} nodes. (The actual samples include repetitions and are therefore longer than~$|S|$.)
\quad \textbf{(left)}~Node degree distribution. Error metric is the Kolmogorov-Smirnov (KS) statistic, i.e., the max vertical distance between CDFs.
\quad \textbf{(middle)}~Average node degree. %
\quad \textbf{(right)}~Fraction $f\eq 0.5$ of labeled nodes. We select at random fraction $f\eq 0.5$ of all nodes and use MHRW and RWRW to estimate~$f$.
The numbers in the bottom-left corner of each figure indicate how much longer MHRW should be than RWRW to achieve the same error, on average.
}
\label{fig:MHRW_vs_RW_simulations}
\end{figure}

\subsubsection{MHRW avoids high-degree nodes, missing good mixing opportunities}\label{subsec:avoiding hubs}
Finally, in networks that arise in practice, nodes of high degree are more likely to act as hubs, \ie to span different parts of the graph than the low-degree nodes whose connections are usually more local.
Therefore, by avoiding hubs, MHRW misses mixing opportunities. In contrast, RW exploits them.

\begin{figure}
\centering
\subfigure[Tested topology]{
\includegraphics[width=0.15\textwidth]{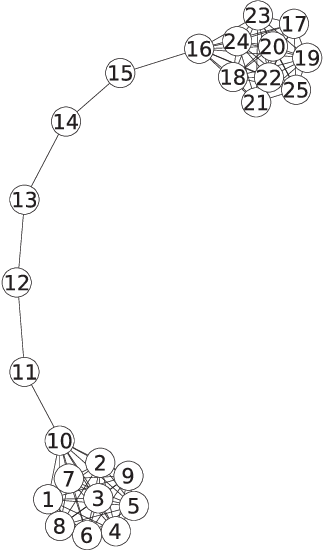}
}
\subfigure[Normalized Root Mean Square Error (NMSE)]{
\includegraphics[width=0.3\textwidth]{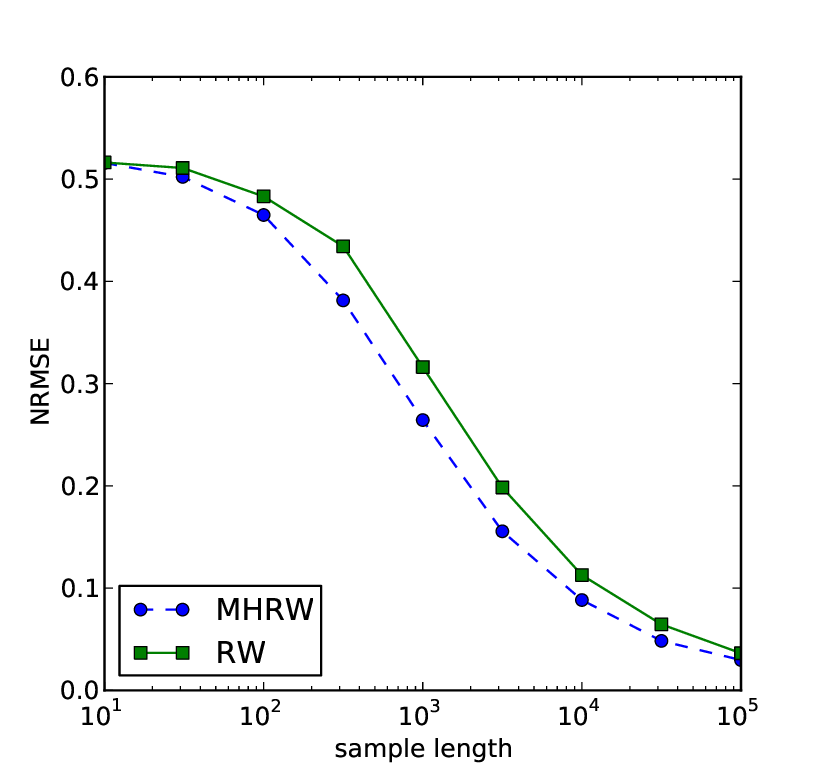}
\label{fig:subfig2}
}
\caption{An example where MHRW performs better than RW. \ \textbf{(a)}~Tested toy graph. It consists of two cliques connected by a 6-hop path. We sample it with RW and MHRW initiated at random nodes. \  \textbf{(b)}~Normalized Root Mean Square Error (NMSE) of the estimation of the average node number (true average is equal to~13), as a function of total sample length. The standard deviation error bars (not shown) are comparable with the symbol size.}
\label{fig:mhrw_counterexample}
\end{figure}

\subsection{Counter-example}

Although RW outperformed MHRW on all real-life topologies we tested, it is not true that RW is always more efficient.
We show one -- carefully constructed and rather unrealistic --
counterexample in Fig.~\ref{fig:mhrw_counterexample}(a). 
Here, every node carries a value equal to its number, and our goal is to estimate the average value $\theta$ (in this example~$\theta=13$).
If the sampling process stays in the clique $v_1\ldots v_{10}$, then the value will be strongly underestimated (probably resulting in $\hat{\theta}\simeq 5.5$. Similarly, staying in the clique $v_{16}\ldots v_{25}$ yields $\hat{\theta}\simeq 20.5$. In order to achieve a good estimate ~$\hat{\theta}\simeq \theta$, our process should be able to switch between the two cliques.

Inside a clique, say inside $v_1\ldots v_{10}$, RW and MHRW behave almost 
 identically. The differences appear once we enter the line section at $v_{11}$. MHRW will typically stay on this line for many iterations (which is not good per se), and eventually end up in one of the cliques with roughly the same probabilities (which is good). In contrast, RW at $v_{11}$ has a much higher chance to return to clique $v_1\ldots v_{10}$, which significantly slows down mixing and thus harms estimation performance.
Similar reasoning applies when leaving clique $v_{16}\ldots v_{25}$ at node~$v_{15}$.
Therefore, MHRW should systematically outperform RW in this example. This intuition is confirmed by the simulation results in~Fig.~\ref{fig:mhrw_counterexample}(b).

\subsection{Summary}

In some pathological cases, like the one presented in Fig.~\ref{fig:mhrw_counterexample},  MHRW may outperform RWRW.
 In more practical examples, at least in all real-life topologies we tried, RWRW requires 1.5-7 times fewer unique samples to achieve the same estimation quality as MHRW, which directly translates into several-fold bandwidth gains. For this reason, we strongly recommend RWRW for sampling nodes in OSN and other Internet topologies.
 However, fully characterizing the relation between graph structure and convergence of MHRW, RWRW is out of the scope of this paper.

\bibliographystyle{IEEEtran}
\bibliography{osn_facebooksampling}

\begin{thebibliography}{10}
\providecommand{\url}[1]{#1}
\csname url@samestyle\endcsname
\providecommand{\newblock}{\relax}
\providecommand{\bibinfo}[2]{#2}
\providecommand{\BIBentrySTDinterwordspacing}{\spaceskip=0pt\relax}
\providecommand{\BIBentryALTinterwordstretchfactor}{4}
\providecommand{\BIBentryALTinterwordspacing}{\spaceskip=\fontdimen2\font plus
\BIBentryALTinterwordstretchfactor\fontdimen3\font minus
  \fontdimen4\font\relax}
\providecommand{\BIBforeignlanguage}[2]{{%
\expandafter\ifx\csname l@#1\endcsname\relax
\typeout{** WARNING: IEEEtran.bst: No hyphenation pattern has been}%
\typeout{** loaded for the language `#1'. Using the pattern for}%
\typeout{** the default language instead.}%
\else
\language=\csname l@#1\endcsname
\fi
#2}}
\providecommand{\BIBdecl}{\relax}
\BIBdecl

\bibitem{nielsen-stats}
``{Nielsen statistics, June 2010},''
  http://blog.nielsen.com/nielsenwire/ %
online\_mobile/social-media-accounts-for-22-percent-of-time-online/ .

\bibitem{alexa}
``{Alexa traffic statistics for Facebook, June 2010},''
  \url{http://www.alexa.com/siteinfo/facebook.com}.

\bibitem{Pujol2010}
J.~Pujol, V.~Erramilli, G.~Siganos, X.~Yang, N.~Laoutaris, P.~Chhabra, and
  P.~Rodriguez, ``{The little engine (s) that could: Scaling online social
  networks},'' \emph{ACM SIGCOMM Computer Communication Review}, vol.~40,
  no.~4, pp. 375--386, 2010.

\bibitem{agarwal2010social}
S.~Agarwal and S.~Agarwal, ``{Social networks as Internet barometers for
  optimizing content delivery networks},'' in \emph{Proc. 3rd IEEE Int. Symp.
  on Advanced Networks and Telecommunication Systems}, New Delhi, India, Dec.
  2009.

\bibitem{nazir2009network}
A.~Nazir, S.~Raza, D.~Gupta, C.~Chuah, and B.~Krishnamurthy, ``{Network level
  footprints of facebook applications},'' in \emph{Proc. 9th ACM SIGCOMM Conf.
  on Internet measurement}, New York, NY, 2009, pp. 63--75.

\bibitem{mislove2008ostra}
A.~Mislove, A.~Post, P.~Druschel, and K.~Gummadi, ``{Ostra: Leveraging trust to
  thwart unwanted communication},'' in \emph{Proc. 5th USENIX Symp. on
  Networked Systems Design and Implementation}, San Francisco, CA, 2008, pp.
  15--30.

\bibitem{Sirivianos-INFOCOM-11}
M.~Sirivianos, X.~Yang, and K.~Kim, ``{SocialFilter: introducing social trust
  to collaborative spam mitigation},'' in \emph{Proc. IEEE INFOCOM}, Shanghai,
  China, 2011.

\bibitem{sirivianos2009facetrust}
M.~Sirivianos, K.~Kim, and X.~Yang, ``{FaceTrust: Assessing the credibility of
  online personas via social networks},'' in \emph{Proc. 4th USENIX Conf. on
  Hot topics in security}, Montreal, Canada, 2009.

\bibitem{Ahn-WWW-07}
Y.~Ahn, S.~Han, H.~Kwak, S.~Moon, and H.~Jeong, ``{Analysis of topological
  characteristics of huge online social networking services},'' in \emph{Proc.
  16th Int. Conf. on World Wide Web}, Banff, Alberta, Canada, 2007, pp.
  835--844.

\bibitem{leskovec2008microscopic}
J.~Leskovec, L.~Backstrom, R.~Kumar, and A.~Tomkins, ``{Microscopic evolution
  of social networks},'' in \emph{Proc. 14th ACM SIGKDD Int. Conf. on Knowledge
  discovery and data mining}, Las Vegas, NV, 2008.

\bibitem{harvard-dataset}
K.~Lewis, J.~Kaufman, M.~Gonzalez, A.~Wimmer, and N.~Christakis, ``{Tastes,
  ties, and time: A new social network dataset using Facebook. com},'' Social
  Networks, pp. 330--342, 2008.

\bibitem{Mislove2007}
A.~Mislove, M.~Marcon, K.~Gummadi, P.~Druschel, and B.~Bhattacharjee,
  ``{Measurement and analysis of online social networks},'' in \emph{Proc. 7th
  ACM SIGCOMM Conf. on Internet measurement}, San Diego, CA, 2007, pp. 29--42.

\bibitem{Wilson09}
C.~Wilson, B.~Boe, A.~Sala, K.~Puttaswamy, and B.~Zhao, ``{User interactions in
  social networks and their implications},'' in \emph{Proc. 4th ACM European
  Conf. on Computer systems}, Nuremberg, Germany, 2009, pp. 205--218.

\bibitem{Kurant2010}
M.~Kurant, A.~Markopoulou, and P.~Thiran, ``{On the bias of BFS (Breadth First
  Search)},'' in \emph{Proc. 22nd Int. Teletraffic Congr., also in
  arXiv:1004.1729}, 2010.

\bibitem{Kurant2011_JSAC_BFS}
------, ``{Towards Unbiased BFS Sampling},'' \emph{To appear in IEEE J. Sel.
  Areas Commun. on Measurement of Internet Topologies}, 2011.

\bibitem{Salganik2004}
M.~Salganik and D.~D. Heckathorn, ``{Sampling and estimation in hidden
  populations using respondent-driven sampling},'' \emph{Sociological
  Methodology}, vol.~34, no.~1, pp. 193--240, 2004.

\bibitem{Rasti09-RDS}
A.~Rasti, M.~Torkjazi, R.~Rejaie, N.~Duffield, W.~Willinger, and D.~Stutzbach,
  ``{Respondent-driven sampling for characterizing unstructured overlays},'' in
  \emph{Proc. IEEE INFOCOM Mini-conference}, Rio de Janeiro, Brazil, 2009, pp.
  2701--2705.

\bibitem{Stutzbach2006-unbiased-p2p}
D.~Stutzbach, R.~Rejaie, N.~Duffield, S.~Sen, and W.~Willinger, ``{On unbiased
  sampling for unstructured peer-to-peer networks},'' in \emph{Proc. 6th ACM
  SIGCOMM Conf. on Internet measurement}, Rio de Janeiro, Brazil, 2006.

\bibitem{Rasti2008}
A.~H. Rasti, M.~Torkjazi, R.~Rejaie, and D.~Stutzbach, ``{Evaluating Sampling
  Techniques for Large Dynamic Graphs},'' \emph{Univ. Oregon, Tech. Rep.
  CIS-TR-08-01}, Sep. 2008.

\bibitem{Twitter08}
B.~Krishnamurthy, P.~Gill, and M.~Arlitt, ``{A few chirps about twitter},'' in
  \emph{Proc. 1st workshop on Online social networks}, Seattle, WA, 2008, pp.
  19--24.

\bibitem{our-dataset}
``{Uniform Sampling of Facebook Users: Publicly Available Datasets, 2009},''
  \url{http://odysseas.calit2.uci.edu/osn}.

\bibitem{wasserman.faust}
S.~Wasserman and K.~Faust, \emph{{Social Network Analysis: Methods and
  Applications}}.\hskip 1em plus 0.5em minus 0.4em\relax Cambridge University
  Press, 1994.

\bibitem{MisloveWosn08}
A.~Mislove, H.~Koppula, K.~Gummadi, P.~Druschel, and B.~Bhattacharjee,
  ``{Growth of the flickr social network},'' in \emph{Proc. 1st workshop on
  Online social networks}, Seattle, WA, 2008, pp. 25--30.

\bibitem{Viswanath2009}
B.~Viswanath, A.~Mislove, M.~Cha, and K.~Gummadi, ``{On the evolution of user
  interaction in facebook},'' in \emph{Proc. 2nd workshop on Online social
  networks}, Barcelona, Spain, 2009, pp. 37--42.

\bibitem{Lee-Phys-Rev-06}
S.~H. Lee, P.-J. Kim, and H.~Jeong, ``{Statistical properties of sampled
  networks},'' \emph{Physical Review E}, vol.~73, p. 16102, 2006.

\bibitem{snowball-bias}
L.~Becchetti, C.~Castillo, D.~Donato, A.~Fazzone, and I.~Rome, ``{A comparison
  of sampling techniques for web graph characterization},'' in \emph{Proc.
  Workshop on Link Analysis}, Philadelphia, PA, 2006.

\bibitem{Ye2010}
S.~Ye, J.~Lang, and F.~Wu, ``{Crawling online social graphs},'' in \emph{Proc.
  12th Asia-Pacific Web Conference}, Busan, Korea, 2010, pp. 236--242.

\bibitem{Lovasz93}
L.~Lov\'{a}sz, ``{Random walks on graphs: A survey},'' \emph{Combinatorics,
  Paul Erdos is Eighty}, vol.~2, no.~1, pp. 1--46, 1993.

\bibitem{Henzinger2000}
M.~R. Henzinger, A.~Heydon, M.~Mitzenmacher, and M.~Najork, ``{On near-uniform
  URL sampling},'' in \emph{Proc. 9th Int. Conf. on World Wide Web}, Amsterdam,
  Netherlands, 2000.

\bibitem{comparison-samplingweb}
E.~Baykan, M.~Henzinger, S.~Keller, S.~{De Castelberg}, and M.~Kinzler, ``{A
  comparison of techniques for sampling web pages},'' in \emph{Proc. 26th Int.
  Symp. on Theoretical Aspects of Computer Science}, Freiburg, Germany, 2009.

\bibitem{Gkantsidis2004}
C.~Gkantsidis, M.~Mihail, and A.~Saberi, ``{Random walks in peer-to-peer
  networks},'' in \emph{Proc. IEEE INFOCOM}, Hong Kong, China, 2004.

\bibitem{Leskovec2006_sampling_from_large_graphs}
J.~Leskovec and C.~Faloutsos, ``{Sampling from large graphs},'' in \emph{Proc.
  12th ACM SIGKDD Int. Conf. on Knowledge discovery and data mining},
  Philadelphia, PA, 2006, pp. 631--636.

\bibitem{Avrachenkov2010}
K.~Avrachenkov, B.~Ribeiro, and D.~Towsley, ``{Improving Random Walk Estimation
  Accuracy with Uniform Restarts},'' in \emph{I7th Workshop on Algorithms and
  Models for the Web Graph}, 2010.

\bibitem{Ribeiro2010}
B.~Ribeiro and D.~Towsley, ``{Estimating and sampling graphs with
  multidimensional random walks},'' in \emph{Proc. 10th ACM SIGCOMM Conf. on
  Internet measurement}, Melbourne, Australia, 2010.

\bibitem{Kurant2011_SWRW}
M.~Kurant, M.~Gjoka, C.~T. Butts, and A.~Markopoulou, ``{Walking on a Graph
  with a Magnifying Glass: Stratified Sampling via Weighted Random Walks},'' in
  \emph{Sigmetrics}, 2011.

\bibitem{Gjoka2011_multigraph_JSAC}
M.~Gjoka, C.~T. Butts, M.~Kurant, and A.~Markopoulou, ``{Multigraph Sampling of
  Online Social Networks},'' \emph{To appear in IEEE J. Sel. Areas Commun. on
  Measurement of Internet Topologies}, 2011.

\bibitem{Latapy2008}
M.~Latapy and C.~Magnien, ``{Complex network measurements: Estimating the
  relevance of observed properties},'' in \emph{Proc. IEEE INFOCOM}, Phoenix,
  AZ, 2008, pp. 1660--1668.

\bibitem{Kumar-KDD-06}
R.~Kumar, J.~Novak, and A.~Tomkins, ``{Structure and Evolution of Online Social
  Networks},'' in \emph{Proc. 12th ACM SIGKDD Int. Conf. on Knowledge discovery
  and data mining}, Philadelphia, PA, 2006.

\bibitem{Backstrom-KDD-06}
L.~Backstrom, D.~Huttenlocher, J.~Kleinberg, and X.~Lan, ``{Group Formation in
  Large Social Networks: Membership, Growth, and Evolution},'' in \emph{Proc.
  12th ACM SIGKDD Int. Conf. on Knowledge discovery and data mining},
  Philadelphia, PA, 2006.

\bibitem{sarkar}
P.~Sarkar and A.~W. Moore, ``{Dynamic social network analysis using latent
  space models},'' \emph{ACM SIGKDD Explorations Newsletter}, vol.~7, no.~2,
  pp. 31--40, Dec. 2005.

\bibitem{Willinger09-OSN_Research}
W.~Willinger, R.~Rejaie, M.~Torkjazi, M.~Valafar, and M.~Maggioni, ``{OSN
  Research: Time to face the real challenges},'' in \emph{Proc. of 2nd Workshop
  on Hot Topics in Measurement \& Modeling of Computer Systems}, Seattle, WA,
  2009.

\bibitem{Kolaczyk2009}
E.~D. Kolaczyk, ``{Statistical Analysis of Network Data},'' \emph{Springer
  Series in Statistics}, vol.~69, no.~4, 2009.

\bibitem{Stumpf2005}
M.~Stumpf, C.~Wiuf, and R.~May, ``{Subnets of scale-free networks are not
  scale-free: sampling properties of networks},'' \emph{Proc. of the Nat.
  Academy of Sciences of the United States of America}, vol. 102, no.~12, p.
  4221, Mar. 2005.

\bibitem{krishnamurthy2009measure}
B.~Krishnamurthy, ``{A measure of online social networks},'' in \emph{Proc. 1st
  Int. Conf. on Communication Systems and Networks}, Bangalore, India, Jan.
  2009.

\bibitem{caltech-dataset}
M.~A. Porter, ``{Facebook5 data},''
  \url{http://www.insna.org/software/data.html}, 2008.

\bibitem{Bala08Privacy}
B.~Krishnamurthy and C.~Wills, ``{Characterizing privacy in online social
  networks},'' in \emph{Proc. 1st workshop on Online social networks}, Seattle,
  WA, 2008, pp. 37--42.

\bibitem{Bonneau2009}
J.~Bonneau, J.~Anderson, R.~Anderson, and F.~Stajano, ``{Eight friends are
  enough: Social graph approximation via public listings},'' in \emph{Proc.
  EuroSys Workshop on Social Network Systems}, Nuremberg, Germany, 2009, pp.
  13--18.

\bibitem{pokingfacebook}
M.~Gjoka, M.~Sirivianos, A.~Markopoulou, and X.~Yang, ``{Poking Facebook:
  characterization of OSN applications},'' in \emph{Proc. 1st workshop on
  Online social networks}, Seattle, WA, 2008, pp. 31--36.

\bibitem{Liben-Nowell-PNAS-05}
D.~Liben-Nowell, J.~Novak, R.~Kumar, P.~Raghavan, and A.~Tomkins, ``{Geographic
  routing in social networks},'' \emph{Proc. of the Nat. Academy of Sciences of
  the United States of America}, vol. 102, no.~33, p. 11623, 2005.

\bibitem{Cha-IMC-07}
M.~Cha, H.~Kwak, P.~Rodriguez, Y.-Y. Ahn, and S.~Moon, ``{I Tube, You Tube,
  Everybody Tubes: Analyzing the World's Largest User Generated Content Video
  System},'' in \emph{Proc. 7th ACM SIGCOMM Conf. on Internet measurement}, San
  Diego, CA, 2007.

\bibitem{youtube}
P.~Gill, M.~Arlitt, Z.~Li, and A.~Mahanti, ``{Youtube traffic characterization:
  a view from the edge},'' in \emph{Proc. 7th ACM SIGCOMM Conf. on Internet
  measurement}, San Diego, CA, 2007.

\bibitem{Gjoka2010}
M.~Gjoka, M.~Kurant, C.~T. Butts, and A.~Markopoulou, ``{Walking in Facebook: A
  Case Study of Unbiased Sampling of OSNs},'' in \emph{Proc. IEEE INFOCOM}, San
  Diego, CA, 2010.

\bibitem{Ribeiro2010a}
B.~Ribeiro, P.~Wang, and D.~Towsley, ``{On Estimating Degree Distributions of
  Directed Graphs through Sampling},'' \emph{UMass, Tech. Rep. UM-CS-2010-046},
  2010.

\bibitem{haupt2008compressed}
J.~Haupt, W.~Bajwa, M.~Rabbat, and R.~Nowak, ``{Compressed sensing for
  networked data},'' \emph{IEEE Signal Processing Mag.}, vol.~25, no.~2, pp.
  92--101, 2008.

\bibitem{kurant11_coarsetopology}
M.~Kurant, M.~Gjoka, Y.~Wang, Z.~W. Almquist, C.~T. Butts, and A.~Markopoulou,
  ``{Coarse-Grained Topology Estimation via Graph Sampling},'' \emph{Arxiv
  preprint arXiv:1105.5488}, 2011.

\bibitem{HansenHurwitz1943}
M.~Hansen and W.~Hurwitz, ``{On the Theory of Sampling from Finite
  Populations},'' \emph{Annals of Mathematical Statistics}, vol.~14, no.~3,
  1943.

\bibitem{VolzHeckathorn08}
E.~Volz and D.~D. Heckathorn, ``{Probability based estimation theory for
  respondent driven sampling},'' \emph{Journal of Official Statistics},
  vol.~24, no.~1, pp. 79--97, 2008.

\bibitem{Metropolis1953}
N.~Metropolis, A.~W. Rosenbluth, M.~N. Rosenbluth, A.~H. Teller, and E.~Teller,
  ``{Equation of state calculation by fast computing machines},'' \emph{Journal
  of Chemical Physics}, vol.~21, pp. 1087--1092, 1953.

\bibitem{mcmc-book}
W.~R. Gilks, S.~Richardson, and D.~J. Spiegelhalter, \emph{{Markov Chain Monte
  Carlo in Practice}}.\hskip 1em plus 0.5em minus 0.4em\relax Chapman and
  Hall/CRC, 1996.

\bibitem{leon-garcia}
A.~Leon-Garcia, \emph{{Probability, statistics, and random processes for
  electrical engineering}}.\hskip 1em plus 0.5em minus 0.4em\relax
  Pearson/Prentice Hall, 2008.

\bibitem{userid64bit}
``{Facebook announcement on 64 bit userIDs, May 2009},''
  \url{http://developers.facebook.com/blog/post/226}.

\bibitem{geweke}
J.~Geweke, ``{Evaluating the accuracy of sampling-based approaches to the
  calculation of posterior moments},'' in \emph{Bayesian Statistics}, 1992, pp.
  169--193.

\bibitem{gelman-rubin}
A.~Gelman and D.~Rubin, ``{Inference from iterative simulation using multiple
  sequences},'' in \emph{Statistical science}, vol.~7, no.~4, 1992, pp.
  457--472.

\bibitem{fb-regional}
``{The facebook blog: Growing beyond regional networks, June 2009},''
  \url{http://blog.facebook.com/blog.php?post=91242982130}.

\bibitem{facebook_uidspacehistory}
``{Facebook's userID numbering system},'' \url{http://www.quora.com/
  What-is-the-history-of-Facebooks-ID-numbering-system}.

\bibitem{Rubin:SIR}
D.~Rubin, ``{Using the SIR algorithm to simulate posterior distributions},'' in
  \emph{Bayesian Statistics}, 1988, vol.~3, pp. 395--402.

\bibitem{Skare03}
O.~Skare, E.~B\o~lviken, and L.~Holden, ``{Improved Sampling-Importance
  Resampling and Reduced Bias Importance Sampling},'' in \emph{Scandinavian
  Journal of Statistics}, 2003, vol.~30, no.~4, pp. 719--737.

\bibitem{Newman02}
M.~Newman, ``{Assortative mixing in networks},'' \emph{Physical Review
  Letters}, vol.~89, no.~20, p. 208701, 2002.

\bibitem{facebook-data}
``{Facebook statistics, Dec 2010},''
  \url{http://facebook.com/press/info.php?statistics}.

\bibitem{inside-bacebook}
``{Inside facebook, July 2009},''
  http://www.insidefacebook.com/2009/07/02/ %
 facebook-now-growing-by-over-700000-users-a-day-updated-engagement-stats/ .

\bibitem{WWW_SNAP_Graph_Library}
``{SNAP Graph Library},'' \url{http://snap.stanford.edu/data/}.

\bibitem{Leskovec2007}
J.~Leskovec, J.~Kleinberg, and C.~Faloutsos, ``{Graph evolution: Densification
  and shrinking diameters},'' \emph{ACM Transactions on Knowledge Discovery
  from Data}, vol.~1, no.~1, p.~2, Mar. 2007.

\bibitem{Albert-Nature-1999}
R.~Albert, H.~Jeong, and A.~Barab\'{a}si, ``{Diameter of the world-wide web},''
  \emph{Nature}, vol. 401, no. 6749, pp. 130--131, 1999.

\bibitem{Leskovec2009}
J.~Leskovec, K.~Lang, A.~Dasgupta, and M.~Mahoney, ``{Community structure in
  large networks: Natural cluster sizes and the absence of large well-defined
  clusters},'' \emph{Internet Mathematics}, vol.~6, no.~1, pp. 29--123, 2009.

\end{thebibliography}

\vfill

\balance

\balance

\end{document}